\documentclass[aps,prb,nobalancelastpage,superscriptaddress,twocolumn,longbibliography]{revtex4-1}
\usepackage[utf8]{inputenc}
\usepackage[american,british]{babel}
\usepackage[T1]{fontenc}
\usepackage[pdftex]{graphicx}  
\usepackage{graphicx, xcolor}
\usepackage{dcolumn}
\usepackage{bm}
\usepackage{amsmath,amsthm,amssymb,amsfonts,mathtools}
\usepackage{bbold}
\usepackage{color}
\usepackage{verbatim}
\usepackage[normalem]{ulem}
\usepackage{soul}

\usepackage{hyperref}
\hypersetup{
 colorlinks=true,
 linkcolor=blue,
 anchorcolor = blue,
 citecolor = blue,
 filecolor = blue,
 urlcolor = blue 
}

\newcommand{\braket}[1]{\langle #1 \rangle}

\newcommand{\notpropto}{\mathrel{\vcenter{
			\offinterlineskip\halign{\hfil$##$\cr
				\propto\cr\noalign{\kern-8.pt}  \rotatebox[origin=c]{-15}{/}\cr\noalign{\kern-2.5pt}}}}}

\newcommand{\nodagger}{{\vphantom{\dagger}}}

\newcommand{\mytitle}{
Spin fractionalization at the edge of quantum Hall fluids induced by bulk quasiparticles}

\begin{document}
 
\title{\mytitle}      

\author{Alexander Fagerlund}
\email{alexander.fagerlund@fysik.su.se}
\affiliation{Department of Physics, Stockholm University, AlbaNova University Center, 106 91 Stockholm, Sweden}

\author{Alberto Nardin}
\email{alberto.nardin@universite-paris-saclay.fr}
\affiliation{Universit\'e Paris-Saclay, CNRS, LPTMS, 91405 Orsay, France}
 
\author{Leonardo Mazza}
\email{leonardo.mazza@universite-paris-saclay.fr}
\affiliation{Universit\'e Paris-Saclay, CNRS, LPTMS, 91405 Orsay, France}
\affiliation{Institut Universitaire de France, 75005 Paris, France}

\author{Eddy Ardonne}
\email{ardonne@fysik.su.se}
\affiliation{Department of Physics, Stockholm University, AlbaNova University Center, 106 91 Stockholm, Sweden}

\begin{abstract}
We define a measurable spin for the edge of a lowest Landau level and incompressible fractional quantum Hall state in the presence of an Abelian or non-Abelian bulk quasiparticle.
We show that this quantity takes a fractional value inherited from the fractional spin of the bulk quasiparticle.
We present a geometric picture that does not rely on global symmetries of the wavefunction but is able to treat quasiparticles and edges with different shapes.
We study finite-size many-body wavefunctions on the cylinder with circular quasiparticles and straight edges.
Our results are supported by matrix-product-state calculations for the Laughlin and the $k=3$ Read-Rezayi states.
\end{abstract}

\maketitle
 
\section{Introduction} 
Fractionalization, namely the fact that the emergent quasiparticles of a many-body setup cannot be interpreted as simple combinations of its elementary constituents,
plays a key role in topological condensed-matter physics 
and constitutes one of its most intriguing phenomenology.
Solitons in one-dimensional polyacetylene molecules~\cite{ssh_prl1979}, spin-1/2 boundary modes in one-dimensional spin-1 chains~\cite{aklt_1987},
zero-energy Majorana modes~\cite{Kitaev_2001}, to mention just a few, are all characterized by some form of fractionalization.

Another celebrated form of fractionalization is that of the bulk quasiparticles of the fractional quantum Hall (FQH) effect~\cite{Feldman_2021}: 
they fractionalize charge~\cite{laughlin_1983}, statistics~\cite{halperin_1984, aarovas_1984} and spin~\cite{Li_1992, Sondhi_1992, Einarsson_1995, Comparin_2022, nardin_2023, Trung_2023}.
Remarkably, the bulk-boundary correspondence dictates a close relation between the fractional properties of the quasiparticles and those of the edge modes; in the case of charge and statistics, the latter have been revealed by celebrated experiments~\cite{de_Picciotto_1997, Saminadayar_1997, Reznikov_1999, Bartolomei_2020, Nakamura_2020, Willett_2023, Glidic_2023, Ruelle_2023, Veillon_2024}.
Whereas other fractionalization properties have prompted research based on effective boundary field theories~\cite{Snizhko_2015, Schiller_2022, Schiller_2024} or entanglement structures~\cite{Tu_2013, Zaletel_2013},
a definition of an {\it edge spin} based solely on measurable edge properties of many-body wavefunctions is lacking.  
 
In this paper we study the spin fractionalization between one bulk quasiparticle and the edge.
Whereas the fractionalization of charge between quasiparticles and edge is ultimately protected by the fundamental charge-conservation law and by the fact that the bulk 
of an incompressible FQH state enjoys the screening hypothesis,
a similarly simple analysis cannot be performed for spin.
In fact the only definition of an edge spin prior to this work was given in Ref.~[\onlinecite{nardin_2023}] for the non-generic case of a circularly symmetric edge.
Fractionalization in the FQH effect is ultimately a topological property and an example of the celebrated bulk-boundary correspondence; it should not depend on the fine-tuned geometric details of the boundary, and in particular on whether it has a circular shape.

In order to prove the aforementioned spin fractionalization in a generic setting, we define a measurable edge spin for incompressible FQH fluids, generalizing that for bulk quasiparticles~\cite{Sondhi_1992,  Comparin_2022}.
We focus on the cylinder geometry, where the straight edge and the circular quasiparticle do not have the same shape.
This forces us to develop a geometric picture of the boundary and to highlight the genuine topological nature of spin fractionalization without relying on global symmetries; 
the resulting notion of spin is disconnected from the intuitive idea of circular rotation.
We present matrix-product-state (MPS) calculations~\cite{Zaletel_2012} for the Laughlin~\cite{laughlin_1983} and the Read-Rezayi (RR) $k=3$~\cite{Read_1999} states that are detailed in an accompanying paper~\cite{Fagerlund_2024} (see Refs.~\cite{Estienne_2013, Estienne_2013b, Wu_2014, Wu2015MPS, Herviou_2024} for related approaches). 
We consider both Abelian and non-Abelian quasiholes and characterize their fractional spin as well as that of the edge; the edge spins satisfy a spin-statistics relation reminiscent of that of bulk quasiparticles~\cite{nardin_2023, Trung_2023}.

The paper is organized as follows.
In Sec.~\ref{Sec:II} we introduce the definition of the edge spin for a cylinder and test this definition in the case of a Laughlin state.
We introduce a geometric definition of the spin both of quasiparticles and of the edge in Sec.~\ref{Sec:III}.
The case of Abelian quasiparticles can be studied analytically and is discussed in Sec.~\ref{Sec:IV}.
Sec.~\ref{Sec:V} is devoted to the study of the edge spin induced by non-Abelian bulk quasiparticles; we do that by considering the case of the RR wavefunction.
Finally, our conclusions and a few perspectives are discussed in Sec.~\ref{Sec:VI}.
The article is concluded by two appendices.

Throughout the paper, the plane is pierced by a uniform and orthogonal magnetic field and
the magnetic length is set to unity, $\ell_B =1$.
We only consider incompressible FQH states defined in the lowest Landau level (LLL) and the plane is parametrized by the complex coordinate $z = x+iy$. 
We set $\hbar=1$. 

\section{The edge spin for the cylindrical geometry} 
\label{Sec:II}

In this Section we introduce the definition of the edge spin in the specific case of a cylindrical geometry for a generic incompressible FQH state and employ it to characterize how the edge of a Laughlin state is modified by the presence of a bulk quasihole.

We consider an incompressible LLL FQH state at filling factor $\nu$   on a finite cylinder with circumference $L_y$ and described by the many-body wavefunction $\Psi_0$; assuming translational invariance of the confining potential along $y$, the density profile $\rho_0$ features two straight boundaries.
When a quasiparticle is inserted at $\eta$, deep in the bulk and far from the edges, the new quantum state, $\Psi_1$, has a density profile $\rho_1$ that features two edges that are shifted but still straight, even if the bulk hosts a circular quasiparticle, since the quasiparticle is a screened defect.
The depletion density profile $\delta \rho = \rho_{1} - \rho_0$ can be split into two contributions, 
\begin{equation}
\delta \rho  = \delta \rho_{qp, \eta} + \delta \rho_{e} , 
\label{Eq:Splitting}
\end{equation}
one for the quasiparticle and one for the edges.

\begin{figure*}[t]
 \includegraphics[width=1.4\columnwidth, trim={0.6cm 0.5cm 0.5cm 0.5cm},clip]{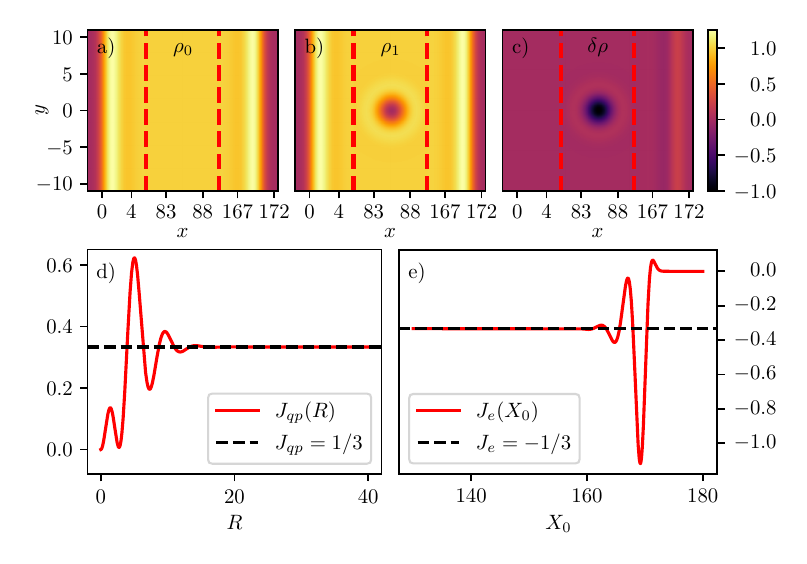}
 \caption{A Laughlin state at $\nu = 1/3$ of $N=200$ fermions on a cylinder with circumference $L_y = 22$. 
 Density profiles in units of $\nu / 2 \pi$: 
 (a) $\rho_0$ without quasiholes; (b) $\rho_1$ with one quasihole in the bulk; (c) $\delta \rho = \rho_{1} - \rho_{0}$; the $x$ axis is broken   (see vertical dashed red lines) to focus on the central part and on the edges.
 Spins:
 (d) The fractional spin of the quasihole is approached in the limit $R \gg 1$ of $J_{qp}(R) = \int_{|z-\eta|\leq R} \left(\frac{|z-\eta|^2}{2}-1 \right) \delta \rho_{qp,\eta} d^2z$ with $\eta= \frac{2 \pi}{L_y} \left(\frac{N}{2 \nu}-1\right) \simeq 85.4$. The numerical curve of $J_{qp}(R)$ (solid red line) approaches a plateau that coincides with the expected value $J_{qp}=1/3$ (dashed black line); (e) The fractional spin of the edge $J_e = -1/3$ (dashed black line) is approached by $J_e(X_0) = \frac{L_y}{2\pi}\int \! d y \int_{X_0}^{+\infty} dx (x-\bar x) \delta \rho_e  $ for $\bar x = \frac{2 \pi}{L_y} \frac{N}{\nu} \simeq 171.4  $ for $X_0 \ll \bar x$.}\label{Fig:Laughlin:Cylinder}
\end{figure*}

We can numerically test these statements on the Laughlin state on the cylinder, which is defined as follows.
Since the Landau-gauge wavefunctions on the cylinder for the LLL can be written as $\phi_{0,q}\propto e^{k_q z} e^{- x^2/2}$, with $z=x+iy$,
the Laughlin wavefunction~\cite{laughlin_1983} at filling $\nu$ can be generalized to the cylinder geometry as~\cite{Thouless_1984}
\begin{equation}
	\Psi_0 \propto \prod_{i<j}(\mu^{z_i} - \mu^{z_j})^{1/\nu} \,e^{-\sum_i \frac{x_i^2}{2}}, \; \, \text{with } \mu=e^{\frac{2\pi}{L_y}}.
\end{equation}
By counting the lowest and highest power of $\mu^{z_i}$, one obtains that the momentum of the lowest occupied orbital corresponds to $k_q=0$, while that of the highest one is $k_q=\frac{2\pi}{L_y}\,\frac{N-1}{\nu}$.

A possible way of inserting $p$ quasiholes at position $\eta$ is given by
$	\Psi_{p} \propto \prod_{i}(\mu^{z_i-\eta} - 1)^p
	\times \Psi_0$;
close to $\eta$, the multiplicative factor is, to first order, $\propto \prod_i (z_i-\eta)^p$, producing a zero of order $p$ in the wavefunction. 
The lowest occupied orbital coincides again with $k_q=0$. On the other hand, the highest occupied one is now $k_q=\frac{2\pi}{L_y}\,\left(\frac{N-1}{\nu}+p\right)$: this choice corresponds to only shifting the right boundary, leaving the left one unaltered even if the quasihole has been inserted.
In this paper, we always study the densest states in the presence of quasiparticles
and restrict ourselves to situations where only the right boundary is modified by the quasiparticle insertion, whereas the left edge is unmodified. This is not the most general situation, but it is conceptually simple and allows us to completely nail down the edge-quasiparticle interplay in the absence of symmetries.
This is the choice that is used throughout the article. 

The validity of the splitting in Eq.~\eqref{Eq:Splitting} is numerically demonstrated in Fig.~\ref{Fig:Laughlin:Cylinder}(a-c) for a Laughlin state in the case of one quasihole~\cite{laughlin_1983}.
The calculation is performed
using an MPS representation of the wavefunctions~\cite{Dubail_2012, Zaletel_2012, Estienne_2013}.
Charge has fractionalized between the boundary and the edge
and it is known since the original paper by Laughlin~\cite{laughlin_1983} that $q_{qp} = \int \! \delta \rho_{qp, \eta} d^2 z$ takes the value $-\nu$ in units of the charge of the elementary constituents of the FQH state.
Since $\int \! \delta \rho \, d^2 z = 0$, it follows that
\begin{equation}
q_{e} = \int \delta \rho_{e} d^2 z= \nu.
\end{equation}

Thanks to the splitting of $\delta \rho$,
a quasiparticle spin 
\begin{equation}
J_{qp} = \int \left(\frac {|z-\eta|^2}{2} - 1 \right) \delta \rho_{qp,\eta} d^2z
\end{equation}
can be defined in terms of the gauge-invariant generator of rotations restricted to the
LLL~\cite{Li_1992, Sondhi_1992, Li_1993, Einarsson_1995, Leinaas2002, Read_2009, Gromov_2016, Umucalilar_2018, Macaluso_2019,  Macaluso_2020, Comparin_2022, Trung_2023}.
In practice, in a finite-size numerical simulation, one needs to integrate over a large region that includes the entire quasiparticle but does not reach the edge. In Fig.~\ref{Fig:Laughlin:Cylinder}(d) we show that for a sufficiently large integration area we obtain the correct value~\cite{Comparin_2022} 
\begin{equation}
J_{qp}= \frac 12 (1-\nu).
\end{equation}
Apart from its clear intrinsic interest, introducing a quasiparticle spin is useful, because it satisfies a spin-statistics relation: the non-local statistics properties of quasiparticles can be obtained via their local spin~\cite{nardin_2023, Trung_2023}.

Since topological phases are characterized by bulk-edge correspondences, it is natural to investigate whether a suitable edge observable inherits the fractional value of $J_{qp}$.
On the cylinder with a straight edge of length $L_y$,
we define the measurable \textit{edge spin} in terms of $\delta \rho_{e}$
\begin{equation}
 J_{e} = \frac{L_y}{2 \pi} \int (x-\bar x) \, \delta \rho_{e} \, d^2z.
 \label{Eq:Spin:Straight:Edge}
\end{equation}
The value of $\bar x$ sets a reference point, or ``center'' of the edge (we discuss how to fix its location below), paralleling the role of $\eta$ in the calculation of the spin of quasiparticles $J_{qp}$. 
We will show that under appropriate conditions $J_e$ takes fractional values related to those of $J_{qp}$, and inherits from it the fact that it satisfies a spin-statistics relation.

Before presenting the reasoning that leads to the definition~\eqref{Eq:Spin:Straight:Edge}, which constitutes the first main result of our work, we proceed to its numerical evaluation in the case of a quasihole in the Laughlin state.
In Fig.~\ref{Fig:Laughlin:Cylinder}(e) we numerically evaluate~\eqref{Eq:Spin:Straight:Edge} by integrating over a large region comprising the entire right boundary but not including the bulk quasihole; we get $J_e = -1/3$ which is exactly the opposite of $J_{qp}$. In Sec.~\ref{Sec:IV} we show that this value can be predicted analytically.

\section{A geometric definition of the spin} \label{Sec:III}
In this Section we propose a
definition of the edge spin in Eq.~\eqref{Eq:Spin:Straight:Edge}, from the viewpoint of the occupation numbers of the orbitals of the LLL. 
We argue that this definition is amenable to extensions to arbitrarily deformed boundaries.
In order to understand what we present in this Section, it is crucial to keep in mind that the LLL is a flat band, for which many bases of orthonormal orbitals are possible; they are equally good for the featureless bulk, but it is important to choose those that have the appropriate shape for the quasiparticles or the edge.

\subsection{The quasiparticle spin}

Consider a generic incompressible FQH state at filling $\nu$.
In the presence of a circular quasiparticle, for instance, it is natural to consider circularly-symmetric orbitals centered at the quasiparticle position, $\eta$.
They are obtained by shifting the symmetric-gauge LLL orbitals and take the form $\phi_{\eta,m} \sim (z-\eta)^m e^{- |z-\eta|^2 /4}$, with $m \in \mathbb N$. 
Their density profile is circular, with average radius $\sqrt{2m}$ and center $\eta$.
We can associate a fermionic second-quantized operator $a^{(\dagger)}_{\eta,m}$ to each of them, and define the orbital occupation numbers  
\begin{equation}
n^{(0)}_{\eta,m} = \langle a^\dagger_{\eta,m} a_{\eta,m} \rangle_{\Psi_0}, 
\qquad
n^{(1)}_{\eta,m} = \langle a^\dagger_{\eta,m} a_{\eta,m} \rangle_{\Psi_1}
\end{equation}
for the states without quasiparticles, $\Psi_0$, and with $1$ quasiparticle $\Psi_1$.
The operator should be fermionic or bosonic depending on the statistics of the elementary constituents of the FQH state; in this paper we only deal with fermions.

As $m$ is increased, the orbital $\phi_{\eta, m}(z)$ explores regions that are further away from the quasiparticle.
For small $m$ and $m'$, $\langle a^\dagger_{\eta, m} a_{\eta, m'} \rangle_{\Psi_i} \approx n^{(i)}_{\eta, m} \delta_{m,m'}$: this is the benefit of having chosen circular orbitals that match the shape of the quasihole
\footnote{We note that due to the screening on the FQH liquid, this relation holds up to corrections that are exponentially small in the distance to the quasiparticle.}%
.
The values of $n^{(0)}_{\eta, m}$ and $n^{(1)}_{\eta, m}$ are different, due to the absence or presence of the quasiparticle. 
For slightly larger values of $m$, the orbitals extend in the bulk of the FQH state, and the occupation numbers $n^{(0)}_{\eta, m}$ and $n^{(1)}_{\eta, m}$ coincide; from general-principle considerations, moreover, they are both equal to $\nu$.
In Appendix~\ref{sec:appB} we show that the spin of the quasiparticle is a function of the orbital occupation numbers:
\begin{equation}
 J_{qp} = \sum_{m=0}^{\Lambda_{m}} m \left( n^{(1)}_{\eta,m}-n^{(0)}_{\eta,m}\right).
 \label{Eq:Spin:qp:orbitals}
\end{equation}
This expression for the quasiparticle spin is inspired by the dipole moment of the edge as studied in Ref.~[\onlinecite{Park_2014}].
The cutoff $\Lambda_{m}$ is introduced to focus only on the quasiparticle  and avoid the edge effects: it should extend up to the intermediate region where the orbitals explore the bulk of the droplet.

For even larger values of $m$, the orbitals reach the edges of the state, and due to the different edge properties of $\Psi_0$ and $\Psi_1$ the occupation numbers will start differing again.
However, differently from what happens close to the quasiparticle,
$\langle a^{\dagger}_{\eta,m} a_{\eta, m'} \rangle_{\Psi_i} \neq 0$ for $m \neq m'$: 
these symmetric-gauge orbitals are not a convenient choice to define the spin of the edge on the cylinder.

\subsection{The edge spin}\label{Sec:III:B}

The natural generalization of $J_{qp}$ to the boundary of the cylinder requires taking the LLL orbitals that mimic the shape of the boundary: in this case these are the shifted Landau-gauge orbitals,
$\phi_{\bar x, q}(z) \sim e^{i k_q y} e^{- (x-x_q - \bar x)^2/2 }$, with $q\in \mathbb Z$, $k_q = 2 \pi q/L_y$ and $x_q = k_q$ (because $\ell_B = 1$);
$\bar x \in 2 \pi \mathbb Z/L_y$ can be chosen  arbitrarily and the orbital is centered around $\bar x+ x_q$. 

By introducing the second-quantized operators $a_{\bar x, q}^{(\dagger)}$ associated to these orbitals, the correlation matrix 
close to the edge is diagonal,
$\langle a_{\bar x, q}^\dagger a_{\bar x, q'} \rangle_{\Psi_i} \propto \delta_{q,q'}$.
Defining the occupation numbers 
\begin{equation}
n^{(i)}_{\bar x, q} = \langle a_{\bar x, q}^\dagger a_{\bar x, q} \rangle_{\Psi_i}
\end{equation}
and taking $\bar x$ close to the boundary, we propose the following spin for the edge, one of our main results:
\begin{equation}
 J_{e} = \sum_{q= -\Lambda_q}^{\infty} q  \left( 
 n^{(1)}_{\bar x,q} -  n^{(0)}_{\bar x,q} \right).
 \label{Eq:Spin:Orb:Straight}
\end{equation}
Close to the edge 
$n^{(0)}_{\bar x, q} \neq n^{(1)}_{\bar x, q}$, but for large  values of $|q|$ they both approach the same values (namely, $0$ for $q \gg 1$ outside of the droplet and $\nu$ for $q \simeq - \Lambda_q$ in the bulk). 
The cutoff $\Lambda_q$ enforces the fact that the orbitals considered in the sum only explore the edge and the neighbouring bulk region, without extending to that part of the bulk where the quasiparticle is located.  

The formulas for $J_{e}$ in Eqs.~\eqref{Eq:Spin:Straight:Edge} and~\eqref{Eq:Spin:Orb:Straight} coincide
whenever $\bar x \in 2 \pi \mathbb Z/L_y$;
the proof is reported below.
Whereas Eq.~\eqref{Eq:Spin:Orb:Straight} is particularly useful in the context of MPS calculations, Eq.~\eqref{Eq:Spin:Straight:Edge} can also be evaluated using Monte-Carlo sampling for simple wavefunctions and can in principle be measured in experiments accessing the density profile of the edge.
By looking at both expressions, we notice that $J_{e}$ has a non-trivial dependence on the parameter $\bar x$.
Since the boundary charge $q_e = \int \delta \rho_ed^2 z $ is equal to $  \sum_{q=-\Lambda_q}^{\infty}(n^{(1)}_{\bar x, q} - n^{(0)}_{\bar x, q})$, a result that is easily proved using the reasoning detailed in Appendix~\ref{sec:appD}, it contributes a term $- \frac{L_y} {2 \pi} q_e \bar x$ to the spin $J_e$. 

We conclude the Section by stressing that
on the cylinder, a bulk quasiparticle and the edge have a different symmetry. Nevertheless, our approach allows us to
define a spin for both, 
disconnecting the notion of spin from that of circular rotation. As we show below, this leads to
the fractionalization of the spin between the quasiparticle and the edge.

\subsection{Proof of the equality of Eqs.~\eqref{Eq:Spin:Straight:Edge} and~\eqref{Eq:Spin:Orb:Straight}}

We consider a straight boundary parallel to the $y$ direction 
and the orthonormal set of straight LLL orbitals $\phi_{\bar x, q} =\sqrt{\frac{1}{L_y \sqrt{\pi}}} e^{i k_q y}e^{- (x- x_q- \bar x)^2/2}$ with $k_q = \frac{2 \pi}{L_y}q$, $x_q = k_q$ and $q \in \mathbb Z$. 
Without loss of generality, we consider the arbitrary value of $\bar x$ to be restricted to the set $2 \pi \mathbb Z / L_y$, chosen in a way to be close to the edge. 
One can easily show that 
\begin{equation}
\frac{L_y}{2 \pi} \int \phi_{\bar x, q'}^* \, x \, \phi_{\bar x, q} dx dy = q\,\delta_{q,q'}. 
\label{Eq:X:P:Landau}
\end{equation}
This equality is stating the  fact that 
the wavefunctions $\phi_{\bar x, q}$ are known to be eigenfunctions of the generator of magnetic translations along $y$, dubbed $T_{y}$, with eigenvalue $e^{i \frac {2 \pi}{L_y}q}$:
when projected to the LLL, the multiplicative operator
$x$ coincides with the LLL projection of $T_{y}$.

Eq.~\eqref{Eq:X:P:Landau} implies the following equality for a generic LLL quantum state $\Psi$
\begin{equation}
    \sum_q q \langle a^\dagger_{\bar x,q} a_{\bar x, q}^\nodagger \rangle_\Psi =
	\frac{L_y}{2 \pi}\int x \rho_{\Psi} dx dy 
	\label{Eq:Straight:State}
\end{equation}
where $\rho_\Psi = \sum_{q,q'} \phi_{\bar x, q'}^*\phi_{\bar x, q}\braket{a^\dagger_{\bar x,q'} a_{\bar x, q}^\nodagger}_{\Psi}$ is the system's density. The $a_{\bar x, q}^{(\dagger)}$ are the second-quantization operators related to the orbitals $\phi_{\bar x, q}$ introduced above.
Using Eq.~\eqref{Eq:Straight:State} the equality between the expressions Eqs.~\eqref{Eq:Spin:Straight:Edge} and~\eqref{Eq:Spin:Orb:Straight} can be shown.

\section{Edge spin in the presence of
Abelian Laughlin quasiholes} 
\label{Sec:IV}

In this Section we show that it is possible to analytically compute the edge spin induced by 
the simplest possible bulk quasiparticles, which are the Abelian Laughlin quasiholes, for a generic incompressible FQH state. 

\subsection{Charge fractionalization induced by the Abelian Laughlin quasihole}

We present the Laughlin quasiholes in the conventional disk notation, even if our numerical study is performed on the cylinder; the local quasiparticle properties are the same. 
They are obtained by inserting an integer number $p$ of magnetic fluxes at position $\eta$ through the ground-state wavefunction $\Psi_0$ of a generic incompressible FQH state at filling $\nu$, obtaining 
\begin{equation}
\Psi_{\mathbb 1,p} \propto \prod_i (z_i-\eta)^p \times \Psi_0.
\end{equation}
Their charges are $q^{(p)}_{qp}=-p \nu$. By charge conservation we have that they induce an excess of charge at the edge, that is,
\begin{equation}
q^{(p)}_e = + p \nu.
\end{equation}

\subsection{Spin of the Abelian Laughlin quasihole for a generic incompressible state}

We now focus on the spin;
before focusing on the edge, we explicitly compute the spin of the Abelian quasihole itself.
The quasiparticle spin can be worked out exploiting circular symmetry. 
When $\Psi_0$ is the Laughlin state it reads~\cite{Comparin_2022, Trung_2023}
\begin{equation}
J^{(p)}_{qp} = - \frac 12 (\nu p^2 -p).
\label{Eq:qp:Spin:1}
\end{equation}
The formula can be generalized to an arbitrary state $\Psi_0$ 
giving
\begin{equation}
J^{(p)}_{qp} = - \frac{1}{2\nu} \left(q^{(p)}_{qp} \right)^2 - \frac{\mathcal S}{2} q^{(p)}_{qp} ,
\label{Eq:qp:Laughlin:spin}
\end{equation}
where $\mathcal S$ is the Wen-Zee shift~\cite{Wen_1992};
indeed, for a Laughlin state, $\mathcal S = \frac 1 \nu$.
The proof is presented here below; the uninterested reader can skip it.

\subsubsection{Proof of Eq.~\eqref{Eq:qp:Laughlin:spin}}

We here prove Eq.~\eqref{Eq:qp:Laughlin:spin} and show the necessity of introducing the Wen-Zee shift $\mathcal S$ for a circularly symmetric Laughlin quasihole by generalizing the analysis carried out in Ref.~[\onlinecite{Comparin_2022}]. 

The total angular momentum of a circularly-symmetric FQH ground-state with $N$ particles can be written as $L_N = \frac{1}{2}N\left(N/\nu-\mathcal{S}\right)$~\cite{haldane2009hallviscosityintrinsicmetric, wavefunctionology, Wen_1992}.
As it is discussed in Appendix~\ref{sec:appB}, the angular momentum can also be computed by integrating the density profile, $\rho_N(\mathbf r)$, multiplied by the multiplicative factor $r^2/2-1$:
\begin{equation}
L_N=\int \rho_N (\mathbf r) 
\left(\frac{r^2}{2}-1 \right) d \mathbf r.
\end{equation}

When a Laughlin quasihole of charge $q_{qh}^{(p)}=-p \nu$ is inserted at $\eta=0$, the angular momentum of the state increases to $L'_N=L_N+pN$.
Crucially, when this quasihole is inserted, the edge of the system is pushed out {\it rigidly}: the density profile $\rho'_N (\mathbf r)$ of the electron gas with the quasihole can be written as the sum of (i) the density profile of an unperturbed state with some extra (fractional) charge $N+\nu p$ and (ii) the density profile of the quasihole. That is, $\rho'_N (\mathbf r) =  \rho_{N+\nu p}(\mathbf r) + \rho_{qp}(\mathbf r)$.

Analogously to the previous case, we can then compute the angular momentum of the state with the quasihole by computing the integral of $\rho'_N(\mathbf r)$ once multiplied by $(r^2 / 2 - 1)$, and match it with the independently known value $L'_N$; we obtain $L'_N = L_{N+p\nu} + J_{qp}^{(p)}$. Solving for $J_{qp}^{(p)}$ gives exactly Eq.~\eqref{Eq:qp:Laughlin:spin}.

\subsection{The edge spin}

The associated edge spin $J^{(p)}_{e}$ can also be worked out, as we now discuss.

\textit{Proposition:} for an Abelian Laughlin quasihole in the Laughlin state of charge $q^{(p)}_{qp}=-\nu p$, the edge spin reads:
\begin{subequations}
\label{Eq:SecondMainResult}
\begin{equation}
 J_{e}^{(p)} = 
\frac{\nu p^2}{2 }- \frac{\nu p}{2} - \frac{L_y}{2 \pi} (\bar x - \bar x_0) \nu p; 
 \label{Eq:Edge:Spin:Analytics}
\end{equation}
when the Abelian Laughlin quasihole is placed in a generic incompressible FQH state, it reads:
\begin{equation}
    J_e^{(p)} =  \frac{1}{2\nu} \left(q^{(p)}_{e} \right)^2 - \frac{ q^{(p)}_{e} }{2} - \frac{L_y}{2 \pi} (\bar x - \bar x_0) q_e^{(p)}. 
\end{equation}
In both expressions, $\bar x$ is the reference coordinate, or ``center'' of the edge, that appears in Eq.~\eqref{Eq:Spin:Straight:Edge};
on the other hand,
$\bar x_0 \in 2 \pi \mathbb Z/L_y$ is the only coordinate value that satisfies the equation:
\begin{equation}
\sum_{q=-\Lambda_q}^{\infty} \left( n_{\bar x_0,q}^{(0)} - \nu \delta_{q<0} \right) = 0,
\label{Eq:Neutrality}
\end{equation}
\end{subequations}
with $\delta_{q<0} = 1$ for $q<0$ and $0$ otherwise.
The coefficients $\nu \delta_{q<0}$, originally introduced in Ref.~[\onlinecite{Park_2014}], define a box background that needs to be correctly aligned with the occupation numbers $n^{(0)}_{\bar x_0, q}$ via a proper choice of $\bar x_0$ so that the edge is chargeless with respect to this background
\footnote{Conservation of charge implies that one can, in principle, define $\bar{x}_0$ uniquely even in actual physical situations,
where the edge structure depends on details of the experimental setup}.
The proof of this proposition and the associated subtleties are discussed below in the specific subsection~\ref{sec:appE} that concludes the Section. That subsection can be skipped by the uninterested reader; here we rather focus on the physical content of the proposition.

First of all, there is a unique value $\bar x = \bar x_0$ \textit{that only depends on edge properties} such that $J^{(p)}_e$ takes the fractional value 
\begin{subequations}
\begin{equation}
J^{(p)}_e =
\frac 12 \nu p (p-1) 
\end{equation}
when $\Psi_0$ is a Laughlin state,
whereas in the most generic situation of an incompressible FQH state such value reads
\begin{equation}
  J^{(p)}_e =  \frac{1}{2 \nu} \left( q_e^{(e)}\right)^2
    - \frac{1} {2} q_e^{(p)}
\end{equation}
\end{subequations}
Even if these fractions differ from that of the bulk Abelian quasihole, 
given in Eqs.~\eqref{Eq:qp:Spin:1} and~\eqref{Eq:qp:Laughlin:spin},
the part proportional to $p^2$ is exactly opposite, and it is the relevant one for the statistics fractionalization of the quasiparticles~\cite{Thouless_1985, Preskill_2004, nardin_2023}. 
This motivates us to speak of a phenomenon of \textit{spin fractionalization} between the quasiparticle and the edge, and constitutes the second main result of this paper.

In the plots presented in Fig.~\ref{Fig:Laughlin:Cylinder} for the Laughlin state, we make the choice 
\begin{equation}
-\frac{L_y}{2 \pi} (\bar x - \bar x_0) =
 \frac 12 \left( 1- \frac 1 \nu \right) 
\end{equation}
which is particularly interesting because it yields the value 
\begin{equation}
J_e^{(p)} = - J_{qp}^{(p)}
\label{Eq:Spin:Fractionalization:1}
\end{equation}
and makes the bulk-edge spin fractionalization phenomenon particularly visible. 

As a second remark, we observe that analogously to the quasiparticle spins, these edge spins satisfy a spin-statistics relation~\cite{nardin_2023}, regardless of the choice of $\bar x$.
One can single out the spin part proportional to $p^2$ by considering the difference $J_e^{(2p)} - 2 J_e^{(p)} $, in a way that is reminiscent of the spin-statistics relation proposed for bulk quasiholes~\cite{Comparin_2022, nardin_2023}.
It is important to stress that $J_e^{(p)}$ does not depend on whether the quasihole in the bulk has broken up into $p$ smaller quasiholes of charge $- \nu$ nor on their circular symmetry: as long as the quasiholes remain far from the edge, the FQH screening protects the value of $J_e^{(p)}$, showing how the spin value is robustly encoded into the boundary.

\subsection{Proof of the Proposition in Eqs.~\eqref{Eq:SecondMainResult}}
\label{sec:appE}

We follow Park and Haldane~\cite{Park_2014} and introduce a set of occupation numbers
$\bar n^{(0)}_{q} = \nu \delta_{q<0}$ which take the values $\nu$ for $ q < 0$ and $0$ for $q \geq 0$.
For the state with a Laughlin Abelian quasihole of charge $q^{(p)}_{qp}$ in the bulk, 
we consider instead the rigidly-shifted orbital occupation numbers 
$\bar n_{q}^{(p)} = \nu \delta_{q < p}$. 
They do not represent a physical state but are a useful computational tool.

First of all, we can write:
\begin{align}
 J_e^{(p)} = \sum_{q=-\Lambda_q}^\infty q \left(n^{(p)}_{\bar x,q} - n^{(0)}_{\bar x, q} \right) = \bar J_e^{(p)} + \Delta_e^{(p)},
\end{align}
with the definitions
\begin{equation}
 \bar J_e^{(p)} = \sum_{q=-\Lambda_q}^\infty q \left(\bar n^{(p)}_{q} - \bar n^{(0)}_{ q} \right) = \frac{\nu p (p-1)}{2}, \label{Eq:EM:barJe}
\end{equation}
which was easily evaluated, 
and 
\begin{equation}
 \Delta_e^{(p)} = \sum_{q=-\Lambda_q}^\infty q 
 \left(n^{(p)}_{\bar x,q} - \bar n^{(p)}_{q} \right)-
 \sum_{q=-\Lambda_q}^\infty q \left(n^{(0)}_{\bar x,q} - \bar n^{(0)}_{ q}\right).
\end{equation}
For the evaluation of $\Delta_e^{(p)}$, we observe that $\bar n^{(p)}_q = \bar n^{(0)}_{q-p}$. Furthermore, a similar relation holds also for the physical occupation numbers, $ n^{(p)}_{\bar x,q} =  n^{(0)}_{\bar x,q-p}$.
Indeed, the insertion of an Abelian quasihole rigidly shifts the boundary towards larger values of $q$ by exactly $p$ orbitals. 
This fact has been extensively discussed in rotationally symmetric configurations but holds true also in other geometries, as can be verified numerically~\cite{Comparin_2022}.
With this we obtain:
\begin{equation}
 \Delta_e^{(p)} = p\sum_{q=-\Lambda_q}^{\infty} \left(
 n^{(0)}_{\bar x,q} - \bar n^{(0)}_q \right)
\end{equation}
The value of $\Delta^{(p)}_e / p$ depends on $\bar x$ but not on $p$. 
We have already shown in Sec.~\ref{Sec:III:B} that the dependence of $J_e^{(p)}$ on $\bar x$ should be $- \frac{L_y}{2 \pi} q_e^{(p)} \bar x$.
Since, from Eq.~\eqref{Eq:EM:barJe},  $\bar J_e^{(p)}$ does not depend on $\bar x$, these considerations pin down the dependence of $\Delta_e^{(p)}$ on $\bar x$.

We choose the orbitals such that $\sum_{q=-\Lambda_q}^{\infty} n^{(0)}_{\bar x, q} = \nu \mathbb Z$.
This is always possible by shifting the FQH density by the necessary (continuous) amount.
Moreover,
since the bulk orbitals have occupation $\nu$,
this condition does not depend on $\Lambda_q$, but only on the edge.
With this choice, we can show that there is a unique and $p$-independent value of $\bar x $, which we call $ \bar x_0 \in 2\pi\mathbb{Z}/L_y$, such that $\Delta^{(p)}_e = 0$. 

To prove this statement, we notice that we can rewrite 
$\Delta_e^{(p)}/p = \sum_{q=-\Lambda_q}^{\infty} 
n^{(0)}_{\bar x,q}  - \nu \mathbb{Z}$.
If we consider a variation of $\bar x$, $\bar x \rightarrow \bar x+ \Delta\bar x$ with $\Delta\bar x= 2\pi \mathbb{Z}'/L_y$, the ground-state occupation numbers are shifted to $n^{(0)}_{\bar x,q}\rightarrow n^{(0)}_{\bar x,q+\Delta\bar x L_y/(2\pi)}$ due to the structure of the orbitals $\phi_{\bar x,q}$. This gives $\sum_{q=-\Lambda_q}^{\infty} 
n^{(0)}_{\bar x,q} \rightarrow \sum_{q=-\Lambda_q +\Delta\bar x L_y/(2\pi)}^{\infty} 
n^{(0)}_{\bar x,q} = \sum_{q=-\Lambda_q}^{\infty} 
n^{(0)}_{\bar x,q} - \nu \mathbb{Z}'$, since the occupation numbers deep in the bulk are constant and equal to $\nu$. Thus, it is possible to choose a new $\bar x = 2\pi\mathbb{Z}/L_y$ which makes $\Delta_e^{(p)}=0$;
this value is unique because $\Delta_e^{(p)}$ is a linear function of $\bar x$.
As $\Delta_e^{(p)} / p$ does not depend on $p$, $\bar x_0$ does not depend on $p$.


Taken together these observations allow us to conclude that we can always write
\begin{equation}
 \Delta^{(p)}_e = -\frac{L_y}{2 \pi} \left( \bar x - \bar x_0 \right) \nu p,
\end{equation}
which proves the thesis in Eqs.~\eqref{Eq:SecondMainResult}.

\section{Edge spins for the RR state with Abelian and non-Abelian quasiholes} \label{Sec:V}
In this final Section, we numerically investigate the RR $k=3$ state with an Abelian or a non-Abelian quasihole inserted deep in the bulk and discuss the quasiparticle and edge spins.
We start by presenting the RR wavefunction~\cite{Read_1999} in the conventional disk notation, even if our numerical study is performed on the cylinder. We use the formulation put forward in Ref.~[\onlinecite{Cappelli_2001}]:
\begin{equation}
 \Psi_{\rm RR} \sim \;  \mathbb S \left[\varphi_k \right] \times \psi_{M}; \quad \varphi_k =  \prod_{s=1}^k\prod_{i_s<j_s} (z_{i_s}-z_{j_s})^2,
 \label{Eq:RR:Wavefunction}
\end{equation}
where  $\psi_{M}=\prod_{i<j} (z_i - z_j)^M \times e^{- \sum_k |z_k|^2 / 4}$ is the Laughlin wavefunction at filling factor $\nu = \frac 1M$.
The $N$ particles are partitioned into $k$ groups of $N/k$ elements, and the term $\prod_{i_s<j_s}(z_{i_s} - z_{j_s})^2$
describes the correlations within the $s$-th cluster~\cite{Cappelli_2001}.
The symbol $\mathbb S$ denotes the operator symmetrizing over all possible partitions.
The RR state has filling
fraction $\nu = k/(kM+2)$, but in this paper we focus exclusively on the case
$(k,M) = (3,1)$, which is a non-Abelian fermionic state at filling fraction $\nu = 3/5$.

\subsection{Review of results on the quasiholes of the RR state}

Let us briefly review some standard results on the wave functions of the quasiholes of the RR state~\cite{Read_1999,Ardonne_2007}.
The smallest possible quasihole, here denoted $\sigma_1$, is obtained by inserting a zero at position $\eta$ for one of the $k$ groups of constituent particles, and then symmetrizing as
\begin{subequations}
\begin{equation}
	\Psi_{\sigma_1} \sim\; \mathbb S \left[ 
		\prod_{i_1} (z_{i_1} - \eta) 
		\times \varphi_k
		\right] \times \psi_M.
\end{equation}
Its charge is $q_{qp}^{(\sigma_1)}=-1/5$. 
A different kind of quasihole can be constructed inserting an additional zero in a different group, leading to
\begin{equation}
		\Psi_{\sigma_2} \sim \; \mathbb S \left[ 
		\prod_{i_1} (z_{i_1} - \eta) 
		\prod_{i_2} (z_{i_2} - \eta)
		\times \varphi_k
		\right] \times \psi_M.
\end{equation}
This gives rise to the $\sigma_2$ quasihole, which has double the charge of the $\sigma_1$.
In a similar way, one can add an additional zero and retrieve the previously-introduced Abelian Laughlin quasiholes $\mathbb 1$, 
\begin{equation}
 \Psi_{\mathbb 1} \equiv \Psi_{\mathbb 1,p=1} \sim \prod_{i} (z_{i} - \eta) \times \Psi_{\rm RR}, 
\end{equation}
which carries charge $q_{qp}^{(\mathbb 1)} = -3/5$; this latter value coincides with $- \nu$ anticipated before. 
Other quasiparticles that we will consider are (see also Ref.~[\onlinecite{Fagerlund_2024}]): 
(i) the $\psi_1$ quasihole, whose wavefunction reads 
\begin{equation}
		\Psi_{\psi_1} \sim \; \mathbb S \left[ 
		\prod_{i_1} (z_{i_1} - \eta)^2
		\times \varphi_k
		\right] \times \psi_M,
\end{equation}
and
whose charge is $q^{(\psi_1)}_{qp} = - 2/5$;
(ii)
the $\psi_2$ quasihole, whose wavefunction reads 
\begin{equation}
		\Psi_{\psi_2} \sim \; \mathbb S \left[ 
		\prod_{i_1} (z_{i_1} - \eta)^2
		\prod_{i_2} (z_{i_2} - \eta)^2
		\times \varphi_k
		\right] \times \psi_M,
\end{equation} 
and whose charge is $q^{(\psi_2)}_{qp} = - 4/5$;
and finally (iii) the $\varepsilon$ quasihole, whose wavefunction reads 
\begin{equation}
		\Psi_{\varepsilon} \sim \; \mathbb S \left[ 
		\prod_{i_1} (z_{i_1} - \eta)^2
		\prod_{i_2} (z_{i_2} - \eta)
		\times \varphi_k
		\right] \times \psi_M,
\end{equation} 
\end{subequations}
and
whose charge is $q_{qp}^{(\varepsilon)} = - 3/5$.


\subsection{Numerical simulations}\label{sec:appG}
To calculate the properties of the quasiholes of the RR states, we use MPS, an approach that is based on the conformal field theory (CFT) formulation of the state itself.  
The canonical method is in terms of a chiral boson and of the $\mathbb Z_k$ parafermion theory. Here we instead
use three chiral bosons to describe the $k=3$ RR states. There are two reasons for
doing so. Firstly, the chiral boson CFT is easier to deal
with when deriving the MPS description. Secondly, we show that one 
can use MPS to obtain interesting properties of states that can be
described in terms of three chiral boson CFTs, paving the way for other
quantum Hall states of interest that are hard to analyze. We note that two-component
chiral boson MPS studies have been performed~\cite{Crepel_2018,Crepel_2019}
for the Halperin~\cite{Halperin_1983,halperin_1984} and Haldane-Rezayi~\cite{Haldane_1988} states.

In this paper, we do not provide details
of the actual MPS description, which can be found in the accompanying paper authored by two of us~\cite{Fagerlund_2024}.

In the numerical calculations we always place the quasihole in the
center of the system, which itself resides on the cylinder. We denote
the circumference of the cylinder by $L_y$.  Depending
on the quasihole considered, the number of electrons is $N_e = 298$ (for the $\psi_2$), $N_e=299$ (for the $\psi_1$ and $\epsilon$) and $N_e=300$ (for the $\sigma_1$, $\sigma_2$ and $\mathbb{1}$).
Finally, to obtain a finite auxiliary Hilbert space, the highest maximal
angular momentum (that is, the cutoff) of the CFT that we use is $P_{\rm max} = 12$.
When inserting a quasi-hole in the bulk, we again do this in such a way that this only modifies the right edge, while the left edge is unchanged. 
This is the same choice as we made above when discussing the Laughlin state.

\subsection{Charge and spin fractionalization for the Abelian and non-Abelian quasiholes of the RR state}

Charge conservation allows to promptly deduce the charge that fractionalizes at the edge when the quasiparticles are inserted in the bulk: they are simply the opposite: $q_{e}^{(\alpha)} = - q_{qp}^{(\alpha)}$. These theoretical values have been verified numerically \cite{Fagerlund_2024}. A summary of these properties is reported in Table~\ref{Table:RR:Edge:NA}.

\begin{table}[t]
\begin{center}
	\begin{tabular}{c || c | c | c | c | c | c}
		Quasihole type $\alpha$ & $\sigma_1$ & $\sigma_2$ & $\psi_1$ & $\mathbb{1}$ & $\epsilon$ & $\psi_2$ \\
		\hline
		$q_{qp}^{(\alpha)}$ & $-1/5$ & $-2/5$ & $-2/5$ & $-3/5$ & $-3/5$ & $-4/5$ \\
		\hline
		$q_{e}^{(\alpha)}$ & $1/5$ & $2/5$ & $2/5$ & $3/5$ & $3/5$ & $4/5$ \\
		\hline
		$h_{qp}^{(\alpha)}$ & $1/15$ & $1/15$ & $2/3$ & $0$ & $2/5$ & $2/3$ \\
		\hline
		$J_{qp}^{(\alpha)}$ & $1/5$ & $2/5$ & $-1/5$ & $3/5$ & $1/5$ & $0$ \\
		\hline
		$J_{e}^{(\alpha)}$ & $-1/5$ & $-2/5$ & $1/5$ & $-3/5$ & $-1/5$ & $0$ 
	\end{tabular}
\end{center}
 \caption{List of the possible elementary quasiholes in the RR $k=3$ state, together with their edge charges $q^{(\alpha)}_e=-q^{(\alpha)}_{qp}$, scaling dimensions $h^{(\alpha)}_{qp}$ and the theoretical values for the spins fractionalized at the quasiparticle $J^{(\alpha)}_{qp}$ in Eq.~\eqref{eq:generalized_qp_spin} and edge $J^{(\alpha)}_e$ in Eq.~\eqref{Eq:J:fractionalization}.}
 \label{Table:RR:Edge:NA}
\end{table}

\begin{figure}[t]
\includegraphics[width=\columnwidth, trim={0.35cm 0.25cm 0.1cm 0.1cm},clip]{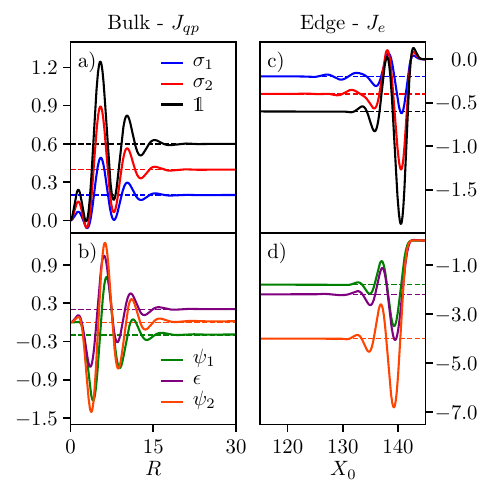}
 \caption{Bulk (a-b) and edge (c-d) spins for the $\nu = 3/5$ RR state with one quasiparticle in the bulk,
 on a cylinder with $L_y = 22$. Calculations are performed in the same way as in Fig.~\ref{Fig:Laughlin:Cylinder}(d-e).
 The different colors label different quasiparticle in the bulk: 
 in the top panels (a,c) the $\sigma_1$ (blue), $\sigma_2$ (red) and $\mathbb{1}$ (black) quasiholes; 
 in the bottom panels (b,d) the $\psi_1$ (green), $\psi_2$ (orange) and $\epsilon$ (purple) quasiholes.
 We used the value $\bar{x} = \frac{2\pi}{Ly}\frac{N}{\nu}\simeq 142.8$, with $N=300$. Dashed lines are the extrapolated values for $J_{qp}^{(\alpha)}$ and $J_e^{(\alpha)}$, corresponding to the values in Table~\ref{Table:RR:Edge:NA} mod~1.
}
 \label{Fig:RR:Edge}
\end{figure}

For a general quasiparticle of type $\alpha$ we propose an expression for its bulk spin $J^{(\alpha)}_{qp}$
which depends on two ground-state properties (the filling factor $\nu$ and the shift $\mathcal S$)
and on two quasiparticle properties (the charge $q_{qp}^{(\alpha)}$ and the coefficients $h_{qp}^{(\alpha)}$ reported in Table~\ref{Table:RR:Edge:NA}),
\begin{equation}
	\label{eq:generalized_qp_spin}
 J_{qp}^{(\alpha_j)} = - \left( 
 \frac{1}{2\nu}
 \left(q_{qp}^{(\alpha_j)} \right)^2 + h_{qp}^{(\alpha_j)} \right) - \frac{\mathcal{S}}{2} q_{qp}^{(\alpha_j)} .
\end{equation}
This form of $J_{qp}^{(\alpha_j)}$ is the natural generalization of the one for Abelian Laughlin quasiholes discussed in Eq.~\eqref{Eq:qp:Laughlin:spin}.
The only difference lies in the first term, which equals minus the scaling dimension of the quasihole,
which has an additional contribution $-h_{qp}^{(\alpha_j)}$. This contribution is due to the non-Abelian nature of the quasihole,
and vanishes for Abelian quasiholes.
Eq.~\eqref{eq:generalized_qp_spin} is numerically verified in Fig.~\ref{Fig:RR:Edge}(a-b), where the numerical evaluation of $J_{qp}^{(\alpha)}$ is compared with the predicted values, that are listed for convenience in Table~\ref{Table:RR:Edge:NA}.

We finally move to the spin of the edge.
Maintaining the previous convention 
\begin{equation}
-\frac{L_y}{2 \pi} (\bar x - \bar x_0) = \frac 12 (1-\mathcal S), 
\label{Eq:Ly:Shift:RR}
\end{equation}
with $\mathcal{S}=M+2$ the shift for the Read-Rezayi series~\cite{Read_1999}, we expect to obtain also in this case the desired fractionalization result:
\begin{equation}
 J_e^{(\alpha)} = - J_{qp}^{(\alpha)}.
 \label{Eq:J:fractionalization}
\end{equation}
Since Ref.~[\onlinecite{nardin_2023}] proved a spin-statistics relation for the $J_{qp}^{(\alpha)}$, Eq.~\eqref{Eq:J:fractionalization} states that the edge-spins too satisfy an analogous one.
The theoretical values for $J_{e}^{(\alpha)}$ obtained via Eq.~\eqref{Eq:J:fractionalization} are reported in Table~\ref{Table:RR:Edge:NA}.

Eq.~\eqref{Eq:J:fractionalization} is numerically verified in Fig.~\ref{Fig:RR:Edge}, by comparing the results in panels (a-b) for the quasiparticles with those in panels (c-d) for the edges.
Notice that in the case of the quasiholes $\psi_1$, $\psi_2$ and $\varepsilon$, the equality is satisfied only modulo $1$.
We trace this fact back to the 
technical subtleties associated with the insertion of the $\psi_1$, $\psi_2$ and $\epsilon$ quasiholes, that require us to compute 
their density profiles 
for different 
particle numbers~\cite{Fagerlund_2024};
note however that the same value of $\bar x$ is employed in the calculations of all the $J_e^{(\alpha)}$.
Because in the spin-statistics-relation setting spins are typically exponentiated as $e^{2 \pi i J}$~\cite{Preskill_2004}, this mismatch is not physically relevant.
The analytical and numerical study of the Abelian and non-Abelian quasiholes of the RR state (in particular the associated bulk and edge spin) constitute the third key result of this paper.

\section{Conclusions} 
\label{Sec:VI}
It is often said that the quasiparticles and the edge of an incompressible FQH state fractionalize charge; in this paper we have presented significant analytical and numerical evidence that \textit{they also fractionalize spin}.
The latter fractionalization is less simple than that of charge, ultimately protected by the fundamental charge-conservation law and the screening property of the bulk of an incompressible FQH state.
Without relying on conservation laws that could be inherited from accidental global symmetries, such as rotational invariance, we have proven the spin fractionalization with general arguments and numerical as well as analytical calculations.

In particular,
we have presented a definition of the edge spin for an incompressible FQH state that is uniquely given in terms of boundary properties, without any explicit reference to the bulk. 
By focusing on the straight edge of a cylinder,
we have shown that the edge spin does not necessarily need to be defined in terms of circular rotations, 
and that a definition can be given both in terms of density integral, Eq.~\eqref{Eq:Spin:Straight:Edge}, or in terms of orbital occupation numbers~\eqref{Eq:Spin:Orb:Straight}.
We have shown analytically and numerically that the spin
fractionalizes between bulk quasiparticles and edge even when the latter has a shape that is not circular, as summarized by Eqs.~\eqref{Eq:Spin:Fractionalization:1} and~\eqref{Eq:J:fractionalization}. 
We have presented numerical MPS simulations for
the cases of the Laughlin and Read-Rezayi states; in the latter case, the quasiparticle and edge spins of the most relevant Abelian and non-Abelian quasiholes have been discussed.

This work contributes to the exciting project of developing a theory of the FQH effect that is intrinsically geometric and does not rely on global symmetries that hide the genuine topological nature of the system~\cite{haldane2009hallviscosityintrinsicmetric, Haldane_2011, Park_2014, haldane2016geometrylandauorbitsabsence, haldane2023incompressiblequantumhallfluids};  it is natural to generalize these notions to the deformed planar droplets presented in Ref.~[\onlinecite{Oblak_2024}].
In particular, it is an intriguing perspective to generalize formulas~\eqref{Eq:Spin:Straight:Edge} and~\eqref{Eq:Spin:Orb:Straight} 
to those geometries.
Moreover, the definition for the quasiparticle spin in terms of orbital occupation numbers given by Eq.~\eqref{Eq:Spin:qp:orbitals} suggests a possible generalization to situations where the quasiparticle is deformed and is not circularly symmetric.


Another intriguing perspective of this study is to understand the experimental consequences of  the fractional edge spin, and how it could be revealed by measurements in state-of-the-art devices. The expression for the edge spin in Eq.~\eqref{Eq:Spin:Orb:Straight}
suggests a possible connection with energy-transport measurements that deserves a deeper analysis.

\acknowledgements

%
We acknowledge insightful discussions with B.~Estienne and B.~Oblak, and also wish to give special thanks to F.~D.~M.~Haldane for illuminating perspectives. 
This work is also part of HQI (www.hqi.fr/en/) initiative and is supported by France 2030 under the French National Research Agency grant number ANR-22-PNCQ-0002.
This research utilized the Sunrise HPC facility supported by the Technical Division at the Department of Physics, Stockholm University.



\appendix

\section{Quasiparticle spin}\label{sec:appB}
We consider a quasiparticle of circular shape centered at $\eta$ and the orthonormal set of concentric circular LLL orbitals $\phi_{\eta, m} = \frac{1}{\sqrt{2 \pi m!}} \left(\frac{z-\eta}{\sqrt 2} \right)^m e^{-|z-\eta|^2/4}$, $m \in \mathbb N$.
One easily shows that
$ \int \phi_{\eta,m'}^* \left( \frac{|z-\eta|^2}{2}-1 \right) \phi_{\eta,m} d^2z=m\,\delta_{m,m'}$. This equality states the following fact. 
The wavefunctions $\phi_{\eta, m}$ are known to be eigenfunctions of the generator of planar rotations around $\eta$, dubbed $L_{\eta}$, with eigenvalue $m$.
When projected to the LLL, the multiplicative operator
$|z-\eta|^2/2-1$ coincides with the LLL projection of $L_{\eta}$~\cite{Comparin_2022}.

This implies the following equality for a generic LLL quantum state $\Psi$
\begin{equation}
    \sum_m m \langle a^\dagger_{\eta,m} a_{\eta, m}^\nodagger \rangle_\Psi =
	\int \left( \frac{|z-\eta|^2}{2}-1 \right) \rho_\Psi d^2z ,
	\label{Eq:Angular:Integral:Orbital}
\end{equation}
where $\rho_\Psi=\sum_{m,m'} \phi_{\eta,m'}^*\phi_{\eta,m}\braket{a_{\eta,m}^{\dagger}a_{\eta,m}^\nodagger}_{\Psi}$ is the system's density. The $a_{\eta,m}^{(\dagger)}$ are the second-quantization operators related to the orbitals $\phi_{\eta,m}$ discussed in the main text.

Using Eq.~\eqref{Eq:Angular:Integral:Orbital} the equality between the two expressions for $J_{qp}$ given in the main text can be shown.

\section{Orbital expression of the charge}\label{sec:appD}

We now rewrite the boundary charge $q_e = \int \delta \rho_e d^2 z$ in terms of occupation numbers.
The integral can be extended over the whole space since $\delta \rho_e$ is exponentially localized at the boundary. As a consequence, using the formulas quoted in the previous paragraph and the orthonormality relation $\int \phi_{\bar x, q'}^*\phi_{\bar x, q} d^2z = \delta_{q,q'}$, we can rewrite the boundary charge as
$q_e =  \sum_{q} (n^{(1)}_{\bar x, q} - n^{(0)}_{\bar x, q})$.
The summation should be understood as being restricted over those orbitals that contribute to $\delta \rho_e$, i.e. those for which $n^{(1)}_{\bar x, q} - n^{(0)}_{\bar x, q}$ is non-zero; namely, $q_e=\sum_{q=-\Lambda_q}^{\infty} (n^{(1)}_{\bar x, q} - n^{(0)}_{\bar x, q})$ for some appropriately chosen cutoff $\Lambda_q$. Notice that the results do not depend on the choice of the cutoff, since deep in the bulk the occupation numbers are identical, $n^{(1)}_{\bar x, q}=n^{(0)}_{\bar x, q}=\nu$.

\bibliography{Edge_Fractionalization_prl_resub.bib}

\begin{thebibliography}{62}%
\makeatletter
\providecommand \@ifxundefined [1]{%
 \@ifx{#1\undefined}
}%
\providecommand \@ifnum [1]{%
 \ifnum #1\expandafter \@firstoftwo
 \else \expandafter \@secondoftwo
 \fi
}%
\providecommand \@ifx [1]{%
 \ifx #1\expandafter \@firstoftwo
 \else \expandafter \@secondoftwo
 \fi
}%
\providecommand \natexlab [1]{#1}%
\providecommand \enquote  [1]{``#1''}%
\providecommand \bibnamefont  [1]{#1}%
\providecommand \bibfnamefont [1]{#1}%
\providecommand \citenamefont [1]{#1}%
\providecommand \href@noop [0]{\@secondoftwo}%
\providecommand \href [0]{\begingroup \@sanitize@url \@href}%
\providecommand \@href[1]{\@@startlink{#1}\@@href}%
\providecommand \@@href[1]{\endgroup#1\@@endlink}%
\providecommand \@sanitize@url [0]{\catcode `\\12\catcode `\$12\catcode `\&12\catcode `\#12\catcode `\^12\catcode `\_12\catcode `\%12\relax}%
\providecommand \@@startlink[1]{}%
\providecommand \@@endlink[0]{}%
\providecommand \url  [0]{\begingroup\@sanitize@url \@url }%
\providecommand \@url [1]{\endgroup\@href {#1}{\urlprefix }}%
\providecommand \urlprefix  [0]{URL }%
\providecommand \Eprint [0]{\href }%
\providecommand \doibase [0]{http://dx.doi.org/}%
\providecommand \selectlanguage [0]{\@gobble}%
\providecommand \bibinfo  [0]{\@secondoftwo}%
\providecommand \bibfield  [0]{\@secondoftwo}%
\providecommand \translation [1]{[#1]}%
\providecommand \BibitemOpen [0]{}%
\providecommand \bibitemStop [0]{}%
\providecommand \bibitemNoStop [0]{.\EOS\space}%
\providecommand \EOS [0]{\spacefactor3000\relax}%
\providecommand \BibitemShut  [1]{\csname bibitem#1\endcsname}%
\let\auto@bib@innerbib\@empty
\bibitem [{\citenamefont {Su}\ \emph {et~al.}(1979)\citenamefont {Su}, \citenamefont {Schrieffer},\ and\ \citenamefont {Heeger}}]{ssh_prl1979}%
  \BibitemOpen
  \bibfield  {author} {\bibinfo {author} {\bibfnamefont {W.~P.}\ \bibnamefont {Su}}, \bibinfo {author} {\bibfnamefont {J.~R.}\ \bibnamefont {Schrieffer}}, \ and\ \bibinfo {author} {\bibfnamefont {A.~J.}\ \bibnamefont {Heeger}},\ }\bibfield  {title} {\enquote {\bibinfo {title} {Solitons in polyacetylene},}\ }\href {\doibase 10.1103/PhysRevLett.42.1698} {\bibfield  {journal} {\bibinfo  {journal} {Phys. Rev. Lett.}\ }\textbf {\bibinfo {volume} {42}},\ \bibinfo {pages} {1698--1701} (\bibinfo {year} {1979})}\BibitemShut {NoStop}%
\bibitem [{\citenamefont {Affleck}\ \emph {et~al.}(1987)\citenamefont {Affleck}, \citenamefont {Kennedy}, \citenamefont {Lieb},\ and\ \citenamefont {Tasaki}}]{aklt_1987}%
  \BibitemOpen
  \bibfield  {author} {\bibinfo {author} {\bibfnamefont {Ian}\ \bibnamefont {Affleck}}, \bibinfo {author} {\bibfnamefont {Tom}\ \bibnamefont {Kennedy}}, \bibinfo {author} {\bibfnamefont {Elliott~H.}\ \bibnamefont {Lieb}}, \ and\ \bibinfo {author} {\bibfnamefont {Hal}\ \bibnamefont {Tasaki}},\ }\bibfield  {title} {\enquote {\bibinfo {title} {Rigorous results on valence-bond ground states in antiferromagnets},}\ }\href {\doibase 10.1103/PhysRevLett.59.799} {\bibfield  {journal} {\bibinfo  {journal} {Phys. Rev. Lett.}\ }\textbf {\bibinfo {volume} {59}},\ \bibinfo {pages} {799--802} (\bibinfo {year} {1987})}\BibitemShut {NoStop}%
\bibitem [{\citenamefont {Kitaev}(2001)}]{Kitaev_2001}%
  \BibitemOpen
  \bibfield  {author} {\bibinfo {author} {\bibfnamefont {A~Yu}\ \bibnamefont {Kitaev}},\ }\bibfield  {title} {\enquote {\bibinfo {title} {Unpaired majorana fermions in quantum wires},}\ }\href {\doibase 10.1070/1063-7869/44/10s/s29} {\bibfield  {journal} {\bibinfo  {journal} {Physics-Uspekhi}\ }\textbf {\bibinfo {volume} {44}},\ \bibinfo {pages} {131–136} (\bibinfo {year} {2001})}\BibitemShut {NoStop}%
\bibitem [{\citenamefont {Feldman}\ and\ \citenamefont {Halperin}(2021)}]{Feldman_2021}%
  \BibitemOpen
  \bibfield  {author} {\bibinfo {author} {\bibfnamefont {D~E}\ \bibnamefont {Feldman}}\ and\ \bibinfo {author} {\bibfnamefont {Bertrand~I}\ \bibnamefont {Halperin}},\ }\bibfield  {title} {\enquote {\bibinfo {title} {Fractional charge and fractional statistics in the quantum hall effects},}\ }\href {\doibase 10.1088/1361-6633/ac03aa} {\bibfield  {journal} {\bibinfo  {journal} {Reports on Progress in Physics}\ }\textbf {\bibinfo {volume} {84}},\ \bibinfo {pages} {076501} (\bibinfo {year} {2021})}\BibitemShut {NoStop}%
\bibitem [{\citenamefont {Laughlin}(1983)}]{laughlin_1983}%
  \BibitemOpen
  \bibfield  {author} {\bibinfo {author} {\bibfnamefont {R.~B.}\ \bibnamefont {Laughlin}},\ }\bibfield  {title} {\enquote {\bibinfo {title} {Anomalous quantum hall effect: An incompressible quantum fluid with fractionally charged excitations},}\ }\href {\doibase 10.1103/PhysRevLett.50.1395} {\bibfield  {journal} {\bibinfo  {journal} {Phys. Rev. Lett.}\ }\textbf {\bibinfo {volume} {50}},\ \bibinfo {pages} {1395--1398} (\bibinfo {year} {1983})}\BibitemShut {NoStop}%
\bibitem [{\citenamefont {Halperin}(1984)}]{halperin_1984}%
  \BibitemOpen
  \bibfield  {author} {\bibinfo {author} {\bibfnamefont {B.~I.}\ \bibnamefont {Halperin}},\ }\bibfield  {title} {\enquote {\bibinfo {title} {Statistics of quasiparticles and the hierarchy of fractional quantized hall states},}\ }\href {\doibase 10.1103/PhysRevLett.52.1583} {\bibfield  {journal} {\bibinfo  {journal} {Phys. Rev. Lett.}\ }\textbf {\bibinfo {volume} {52}},\ \bibinfo {pages} {1583--1586} (\bibinfo {year} {1984})}\BibitemShut {NoStop}%
\bibitem [{\citenamefont {Arovas}\ \emph {et~al.}(1984)\citenamefont {Arovas}, \citenamefont {Schrieffer},\ and\ \citenamefont {Wilczek}}]{aarovas_1984}%
  \BibitemOpen
  \bibfield  {author} {\bibinfo {author} {\bibfnamefont {Daniel}\ \bibnamefont {Arovas}}, \bibinfo {author} {\bibfnamefont {J.~R.}\ \bibnamefont {Schrieffer}}, \ and\ \bibinfo {author} {\bibfnamefont {Frank}\ \bibnamefont {Wilczek}},\ }\bibfield  {title} {\enquote {\bibinfo {title} {Fractional statistics and the quantum hall effect},}\ }\href {\doibase 10.1103/PhysRevLett.53.722} {\bibfield  {journal} {\bibinfo  {journal} {Phys. Rev. Lett.}\ }\textbf {\bibinfo {volume} {53}},\ \bibinfo {pages} {722--723} (\bibinfo {year} {1984})}\BibitemShut {NoStop}%
\bibitem [{\citenamefont {Li}(1992)}]{Li_1992}%
  \BibitemOpen
  \bibfield  {author} {\bibinfo {author} {\bibfnamefont {D.}~\bibnamefont {Li}},\ }\bibfield  {title} {\enquote {\bibinfo {title} {The spin of the quasi-particle in the fractional quantum hall effect},}\ }\href {\doibase https://doi.org/10.1016/0375-9601(92)90810-9} {\bibfield  {journal} {\bibinfo  {journal} {Phys. Lett. A}\ }\textbf {\bibinfo {volume} {169}},\ \bibinfo {pages} {82--86} (\bibinfo {year} {1992})}\BibitemShut {NoStop}%
\bibitem [{\citenamefont {Sondhi}\ and\ \citenamefont {Kivelson}(1992)}]{Sondhi_1992}%
  \BibitemOpen
  \bibfield  {author} {\bibinfo {author} {\bibfnamefont {S.~L.}\ \bibnamefont {Sondhi}}\ and\ \bibinfo {author} {\bibfnamefont {S.~A.}\ \bibnamefont {Kivelson}},\ }\bibfield  {title} {\enquote {\bibinfo {title} {Long-range interactions and the quantum hall effect},}\ }\href {\doibase 10.1103/PhysRevB.46.13319} {\bibfield  {journal} {\bibinfo  {journal} {Phys. Rev. B}\ }\textbf {\bibinfo {volume} {46}},\ \bibinfo {pages} {13319--13325} (\bibinfo {year} {1992})}\BibitemShut {NoStop}%
\bibitem [{\citenamefont {Einarsson}\ \emph {et~al.}(1995)\citenamefont {Einarsson}, \citenamefont {Sondhi}, \citenamefont {Girvin},\ and\ \citenamefont {Arovas}}]{Einarsson_1995}%
  \BibitemOpen
  \bibfield  {author} {\bibinfo {author} {\bibfnamefont {T.}~\bibnamefont {Einarsson}}, \bibinfo {author} {\bibfnamefont {S.~L.}\ \bibnamefont {Sondhi}}, \bibinfo {author} {\bibfnamefont {S.~M.}\ \bibnamefont {Girvin}}, \ and\ \bibinfo {author} {\bibfnamefont {D.~P.}\ \bibnamefont {Arovas}},\ }\bibfield  {title} {\enquote {\bibinfo {title} {Fractional spin for quantum hall effect quasiparticles},}\ }\href {\doibase https://doi.org/10.1016/0550-3213(95)00025-N} {\bibfield  {journal} {\bibinfo  {journal} {Nucl. Phys. B}\ }\textbf {\bibinfo {volume} {441}},\ \bibinfo {pages} {515--529} (\bibinfo {year} {1995})}\BibitemShut {NoStop}%
\bibitem [{\citenamefont {Comparin}\ \emph {et~al.}(2022)\citenamefont {Comparin}, \citenamefont {Opler}, \citenamefont {Macaluso}, \citenamefont {Biella}, \citenamefont {Polychronakos},\ and\ \citenamefont {Mazza}}]{Comparin_2022}%
  \BibitemOpen
  \bibfield  {author} {\bibinfo {author} {\bibfnamefont {Tommaso}\ \bibnamefont {Comparin}}, \bibinfo {author} {\bibfnamefont {Alvin}\ \bibnamefont {Opler}}, \bibinfo {author} {\bibfnamefont {Elia}\ \bibnamefont {Macaluso}}, \bibinfo {author} {\bibfnamefont {Alberto}\ \bibnamefont {Biella}}, \bibinfo {author} {\bibfnamefont {Alexios~P.}\ \bibnamefont {Polychronakos}}, \ and\ \bibinfo {author} {\bibfnamefont {Leonardo}\ \bibnamefont {Mazza}},\ }\bibfield  {title} {\enquote {\bibinfo {title} {Measurable fractional spin for quantum hall quasiparticles on the disk},}\ }\href {\doibase 10.1103/PhysRevB.105.085125} {\bibfield  {journal} {\bibinfo  {journal} {Phys. Rev. B}\ }\textbf {\bibinfo {volume} {105}},\ \bibinfo {pages} {085125} (\bibinfo {year} {2022})}\BibitemShut {NoStop}%
\bibitem [{\citenamefont {Nardin}\ \emph {et~al.}(2023)\citenamefont {Nardin}, \citenamefont {Ardonne},\ and\ \citenamefont {Mazza}}]{nardin_2023}%
  \BibitemOpen
  \bibfield  {author} {\bibinfo {author} {\bibfnamefont {Alberto}\ \bibnamefont {Nardin}}, \bibinfo {author} {\bibfnamefont {Eddy}\ \bibnamefont {Ardonne}}, \ and\ \bibinfo {author} {\bibfnamefont {Leonardo}\ \bibnamefont {Mazza}},\ }\bibfield  {title} {\enquote {\bibinfo {title} {Spin-statistics relation for quantum hall states},}\ }\href {\doibase 10.1103/PhysRevB.108.L041105} {\bibfield  {journal} {\bibinfo  {journal} {Phys. Rev. B}\ }\textbf {\bibinfo {volume} {108}},\ \bibinfo {pages} {L041105} (\bibinfo {year} {2023})}\BibitemShut {NoStop}%
\bibitem [{\citenamefont {Trung}\ \emph {et~al.}(2023)\citenamefont {Trung}, \citenamefont {Wang},\ and\ \citenamefont {Yang}}]{Trung_2023}%
  \BibitemOpen
  \bibfield  {author} {\bibinfo {author} {\bibfnamefont {Ha~Quang}\ \bibnamefont {Trung}}, \bibinfo {author} {\bibfnamefont {Yuzhu}\ \bibnamefont {Wang}}, \ and\ \bibinfo {author} {\bibfnamefont {Bo}~\bibnamefont {Yang}},\ }\bibfield  {title} {\enquote {\bibinfo {title} {Spin-statistics relation and abelian braiding phase for anyons in the fractional quantum hall effect},}\ }\href {\doibase 10.1103/PhysRevB.107.L201301} {\bibfield  {journal} {\bibinfo  {journal} {Phys. Rev. B}\ }\textbf {\bibinfo {volume} {107}},\ \bibinfo {pages} {L201301} (\bibinfo {year} {2023})}\BibitemShut {NoStop}%
\bibitem [{\citenamefont {de~Picciotto}\ \emph {et~al.}(1997)\citenamefont {de~Picciotto}, \citenamefont {Reznikov}, \citenamefont {Heiblum}, \citenamefont {Umansky}, \citenamefont {Bunin},\ and\ \citenamefont {Mahalu}}]{de_Picciotto_1997}%
  \BibitemOpen
  \bibfield  {author} {\bibinfo {author} {\bibfnamefont {R.}~\bibnamefont {de~Picciotto}}, \bibinfo {author} {\bibfnamefont {M.}~\bibnamefont {Reznikov}}, \bibinfo {author} {\bibfnamefont {M.}~\bibnamefont {Heiblum}}, \bibinfo {author} {\bibfnamefont {V.}~\bibnamefont {Umansky}}, \bibinfo {author} {\bibfnamefont {G.}~\bibnamefont {Bunin}}, \ and\ \bibinfo {author} {\bibfnamefont {D.}~\bibnamefont {Mahalu}},\ }\bibfield  {title} {\enquote {\bibinfo {title} {Direct observation of a fractional charge},}\ }\href {\doibase 10.1038/38241} {\bibfield  {journal} {\bibinfo  {journal} {Nature}\ }\textbf {\bibinfo {volume} {389}},\ \bibinfo {pages} {162–164} (\bibinfo {year} {1997})}\BibitemShut {NoStop}%
\bibitem [{\citenamefont {Saminadayar}\ \emph {et~al.}(1997)\citenamefont {Saminadayar}, \citenamefont {Glattli}, \citenamefont {Jin},\ and\ \citenamefont {Etienne}}]{Saminadayar_1997}%
  \BibitemOpen
  \bibfield  {author} {\bibinfo {author} {\bibfnamefont {L.}~\bibnamefont {Saminadayar}}, \bibinfo {author} {\bibfnamefont {D.~C.}\ \bibnamefont {Glattli}}, \bibinfo {author} {\bibfnamefont {Y.}~\bibnamefont {Jin}}, \ and\ \bibinfo {author} {\bibfnamefont {B.}~\bibnamefont {Etienne}},\ }\bibfield  {title} {\enquote {\bibinfo {title} {Observation of the $\mathit{e}\mathit{/}3$ fractionally charged laughlin quasiparticle},}\ }\href {\doibase 10.1103/PhysRevLett.79.2526} {\bibfield  {journal} {\bibinfo  {journal} {Phys. Rev. Lett.}\ }\textbf {\bibinfo {volume} {79}},\ \bibinfo {pages} {2526--2529} (\bibinfo {year} {1997})}\BibitemShut {NoStop}%
\bibitem [{\citenamefont {Reznikov}\ \emph {et~al.}(1999)\citenamefont {Reznikov}, \citenamefont {Picciotto}, \citenamefont {Griffiths}, \citenamefont {Heiblum},\ and\ \citenamefont {Umansky}}]{Reznikov_1999}%
  \BibitemOpen
  \bibfield  {author} {\bibinfo {author} {\bibfnamefont {M.}~\bibnamefont {Reznikov}}, \bibinfo {author} {\bibfnamefont {R.~de}\ \bibnamefont {Picciotto}}, \bibinfo {author} {\bibfnamefont {T.~G.}\ \bibnamefont {Griffiths}}, \bibinfo {author} {\bibfnamefont {M.}~\bibnamefont {Heiblum}}, \ and\ \bibinfo {author} {\bibfnamefont {V.}~\bibnamefont {Umansky}},\ }\bibfield  {title} {\enquote {\bibinfo {title} {Observation of quasiparticles with one-fifth of an electron's charge},}\ }\href {\doibase 10.1038/20384} {\bibfield  {journal} {\bibinfo  {journal} {Nature}\ }\textbf {\bibinfo {volume} {399}},\ \bibinfo {pages} {238--241} (\bibinfo {year} {1999})}\BibitemShut {NoStop}%
\bibitem [{\citenamefont {Bartolomei}\ \emph {et~al.}(2020)\citenamefont {Bartolomei}, \citenamefont {Kumar}, \citenamefont {Bisognin}, \citenamefont {Marguerite}, \citenamefont {Berroir}, \citenamefont {Bocquillon}, \citenamefont {Plaçais}, \citenamefont {Cavanna}, \citenamefont {Dong}, \citenamefont {Gennser}, \citenamefont {Jin},\ and\ \citenamefont {Fève}}]{Bartolomei_2020}%
  \BibitemOpen
  \bibfield  {author} {\bibinfo {author} {\bibfnamefont {H.}~\bibnamefont {Bartolomei}}, \bibinfo {author} {\bibfnamefont {M.}~\bibnamefont {Kumar}}, \bibinfo {author} {\bibfnamefont {R.}~\bibnamefont {Bisognin}}, \bibinfo {author} {\bibfnamefont {A.}~\bibnamefont {Marguerite}}, \bibinfo {author} {\bibfnamefont {J.-M.}\ \bibnamefont {Berroir}}, \bibinfo {author} {\bibfnamefont {E.}~\bibnamefont {Bocquillon}}, \bibinfo {author} {\bibfnamefont {B.}~\bibnamefont {Plaçais}}, \bibinfo {author} {\bibfnamefont {A.}~\bibnamefont {Cavanna}}, \bibinfo {author} {\bibfnamefont {Q.}~\bibnamefont {Dong}}, \bibinfo {author} {\bibfnamefont {U.}~\bibnamefont {Gennser}}, \bibinfo {author} {\bibfnamefont {Y.}~\bibnamefont {Jin}}, \ and\ \bibinfo {author} {\bibfnamefont {G.}~\bibnamefont {Fève}},\ }\bibfield  {title} {\enquote {\bibinfo {title} {Fractional statistics in anyon collisions},}\ }\href {\doibase 10.1126/science.aaz5601} {\bibfield  {journal} {\bibinfo  {journal} {Science}\ }\textbf {\bibinfo {volume} {368}},\
  \bibinfo {pages} {173–177} (\bibinfo {year} {2020})}\BibitemShut {NoStop}%
\bibitem [{\citenamefont {Nakamura}\ \emph {et~al.}(2020)\citenamefont {Nakamura}, \citenamefont {Liang}, \citenamefont {Gardner},\ and\ \citenamefont {Manfra}}]{Nakamura_2020}%
  \BibitemOpen
  \bibfield  {author} {\bibinfo {author} {\bibfnamefont {J.}~\bibnamefont {Nakamura}}, \bibinfo {author} {\bibfnamefont {S.}~\bibnamefont {Liang}}, \bibinfo {author} {\bibfnamefont {G.~C.}\ \bibnamefont {Gardner}}, \ and\ \bibinfo {author} {\bibfnamefont {M.~J.}\ \bibnamefont {Manfra}},\ }\bibfield  {title} {\enquote {\bibinfo {title} {Direct observation of anyonic braiding statistics},}\ }\href {\doibase 10.1038/s41567-020-1019-1} {\bibfield  {journal} {\bibinfo  {journal} {Nature Physics}\ }\textbf {\bibinfo {volume} {16}},\ \bibinfo {pages} {931–936} (\bibinfo {year} {2020})}\BibitemShut {NoStop}%
\bibitem [{\citenamefont {Willett}\ \emph {et~al.}(2023)\citenamefont {Willett}, \citenamefont {Shtengel}, \citenamefont {Nayak}, \citenamefont {Pfeiffer}, \citenamefont {Chung}, \citenamefont {Peabody}, \citenamefont {Baldwin},\ and\ \citenamefont {West}}]{Willett_2023}%
  \BibitemOpen
  \bibfield  {author} {\bibinfo {author} {\bibfnamefont {R.~L.}\ \bibnamefont {Willett}}, \bibinfo {author} {\bibfnamefont {K.}~\bibnamefont {Shtengel}}, \bibinfo {author} {\bibfnamefont {C.}~\bibnamefont {Nayak}}, \bibinfo {author} {\bibfnamefont {L.~N.}\ \bibnamefont {Pfeiffer}}, \bibinfo {author} {\bibfnamefont {Y.~J.}\ \bibnamefont {Chung}}, \bibinfo {author} {\bibfnamefont {M.~L.}\ \bibnamefont {Peabody}}, \bibinfo {author} {\bibfnamefont {K.~W.}\ \bibnamefont {Baldwin}}, \ and\ \bibinfo {author} {\bibfnamefont {K.~W.}\ \bibnamefont {West}},\ }\bibfield  {title} {\enquote {\bibinfo {title} {Interference measurements of non-abelian e/4 and abelian e/2 quasiparticle braiding},}\ }\href {\doibase 10.1103/PhysRevX.13.011028} {\bibfield  {journal} {\bibinfo  {journal} {Phys. Rev. X}\ }\textbf {\bibinfo {volume} {13}},\ \bibinfo {pages} {011028} (\bibinfo {year} {2023})}\BibitemShut {NoStop}%
\bibitem [{\citenamefont {Glidic}\ \emph {et~al.}(2023)\citenamefont {Glidic}, \citenamefont {Maillet}, \citenamefont {Aassime}, \citenamefont {Piquard}, \citenamefont {Cavanna}, \citenamefont {Gennser}, \citenamefont {Jin}, \citenamefont {Anthore},\ and\ \citenamefont {Pierre}}]{Glidic_2023}%
  \BibitemOpen
  \bibfield  {author} {\bibinfo {author} {\bibfnamefont {P.}~\bibnamefont {Glidic}}, \bibinfo {author} {\bibfnamefont {O.}~\bibnamefont {Maillet}}, \bibinfo {author} {\bibfnamefont {A.}~\bibnamefont {Aassime}}, \bibinfo {author} {\bibfnamefont {C.}~\bibnamefont {Piquard}}, \bibinfo {author} {\bibfnamefont {A.}~\bibnamefont {Cavanna}}, \bibinfo {author} {\bibfnamefont {U.}~\bibnamefont {Gennser}}, \bibinfo {author} {\bibfnamefont {Y.}~\bibnamefont {Jin}}, \bibinfo {author} {\bibfnamefont {A.}~\bibnamefont {Anthore}}, \ and\ \bibinfo {author} {\bibfnamefont {F.}~\bibnamefont {Pierre}},\ }\bibfield  {title} {\enquote {\bibinfo {title} {Cross-correlation investigation of anyon statistics in the $\ensuremath{\nu}=1/3$ and $2/5$ fractional quantum hall states},}\ }\href {\doibase 10.1103/PhysRevX.13.011030} {\bibfield  {journal} {\bibinfo  {journal} {Phys. Rev. X}\ }\textbf {\bibinfo {volume} {13}},\ \bibinfo {pages} {011030} (\bibinfo {year} {2023})}\BibitemShut {NoStop}%
\bibitem [{\citenamefont {Ruelle}\ \emph {et~al.}(2023)\citenamefont {Ruelle}, \citenamefont {Frigerio}, \citenamefont {Berroir}, \citenamefont {Pla\ifmmode~\mbox{\c{c}}\else \c{c}\fi{}ais}, \citenamefont {Rech}, \citenamefont {Cavanna}, \citenamefont {Gennser}, \citenamefont {Jin},\ and\ \citenamefont {F\`eve}}]{Ruelle_2023}%
  \BibitemOpen
  \bibfield  {author} {\bibinfo {author} {\bibfnamefont {M.}~\bibnamefont {Ruelle}}, \bibinfo {author} {\bibfnamefont {E.}~\bibnamefont {Frigerio}}, \bibinfo {author} {\bibfnamefont {J.-M.}\ \bibnamefont {Berroir}}, \bibinfo {author} {\bibfnamefont {B.}~\bibnamefont {Pla\ifmmode~\mbox{\c{c}}\else \c{c}\fi{}ais}}, \bibinfo {author} {\bibfnamefont {J.}~\bibnamefont {Rech}}, \bibinfo {author} {\bibfnamefont {A.}~\bibnamefont {Cavanna}}, \bibinfo {author} {\bibfnamefont {U.}~\bibnamefont {Gennser}}, \bibinfo {author} {\bibfnamefont {Y.}~\bibnamefont {Jin}}, \ and\ \bibinfo {author} {\bibfnamefont {G.}~\bibnamefont {F\`eve}},\ }\bibfield  {title} {\enquote {\bibinfo {title} {Comparing fractional quantum hall laughlin and jain topological orders with the anyon collider},}\ }\href {\doibase 10.1103/PhysRevX.13.011031} {\bibfield  {journal} {\bibinfo  {journal} {Phys. Rev. X}\ }\textbf {\bibinfo {volume} {13}},\ \bibinfo {pages} {011031} (\bibinfo {year} {2023})}\BibitemShut {NoStop}%
\bibitem [{\citenamefont {Veillon}\ \emph {et~al.}(2024)\citenamefont {Veillon}, \citenamefont {Piquard}, \citenamefont {Glidic}, \citenamefont {Sato}, \citenamefont {Aassime}, \citenamefont {Cavanna}, \citenamefont {Jin}, \citenamefont {Gennser}, \citenamefont {Anthore},\ and\ \citenamefont {Pierre}}]{Veillon_2024}%
  \BibitemOpen
  \bibfield  {author} {\bibinfo {author} {\bibfnamefont {A.}~\bibnamefont {Veillon}}, \bibinfo {author} {\bibfnamefont {C.}~\bibnamefont {Piquard}}, \bibinfo {author} {\bibfnamefont {P.}~\bibnamefont {Glidic}}, \bibinfo {author} {\bibfnamefont {Y.}~\bibnamefont {Sato}}, \bibinfo {author} {\bibfnamefont {A.}~\bibnamefont {Aassime}}, \bibinfo {author} {\bibfnamefont {A.}~\bibnamefont {Cavanna}}, \bibinfo {author} {\bibfnamefont {Y.}~\bibnamefont {Jin}}, \bibinfo {author} {\bibfnamefont {U.}~\bibnamefont {Gennser}}, \bibinfo {author} {\bibfnamefont {A.}~\bibnamefont {Anthore}}, \ and\ \bibinfo {author} {\bibfnamefont {F.}~\bibnamefont {Pierre}},\ }\bibfield  {title} {\enquote {\bibinfo {title} {Observation of the scaling dimension of fractional quantum hall anyons},}\ }\href {\doibase 10.1038/s41586-024-07727-z} {\bibfield  {journal} {\bibinfo  {journal} {Nature}\ }\textbf {\bibinfo {volume} {632}},\ \bibinfo {pages} {517–521} (\bibinfo {year} {2024})}\BibitemShut {NoStop}%
\bibitem [{\citenamefont {Snizhko}\ and\ \citenamefont {Cheianov}(2015)}]{Snizhko_2015}%
  \BibitemOpen
  \bibfield  {author} {\bibinfo {author} {\bibfnamefont {Kyrylo}\ \bibnamefont {Snizhko}}\ and\ \bibinfo {author} {\bibfnamefont {Vadim}\ \bibnamefont {Cheianov}},\ }\bibfield  {title} {\enquote {\bibinfo {title} {Scaling dimension of quantum hall quasiparticles from tunneling-current noise measurements},}\ }\href {\doibase 10.1103/PhysRevB.91.195151} {\bibfield  {journal} {\bibinfo  {journal} {Phys. Rev. B}\ }\textbf {\bibinfo {volume} {91}},\ \bibinfo {pages} {195151} (\bibinfo {year} {2015})}\BibitemShut {NoStop}%
\bibitem [{\citenamefont {Schiller}\ \emph {et~al.}(2022)\citenamefont {Schiller}, \citenamefont {Oreg},\ and\ \citenamefont {Snizhko}}]{Schiller_2022}%
  \BibitemOpen
  \bibfield  {author} {\bibinfo {author} {\bibfnamefont {Noam}\ \bibnamefont {Schiller}}, \bibinfo {author} {\bibfnamefont {Yuval}\ \bibnamefont {Oreg}}, \ and\ \bibinfo {author} {\bibfnamefont {Kyrylo}\ \bibnamefont {Snizhko}},\ }\bibfield  {title} {\enquote {\bibinfo {title} {Extracting the scaling dimension of quantum hall quasiparticles from current correlations},}\ }\href {\doibase 10.1103/PhysRevB.105.165150} {\bibfield  {journal} {\bibinfo  {journal} {Phys. Rev. B}\ }\textbf {\bibinfo {volume} {105}},\ \bibinfo {pages} {165150} (\bibinfo {year} {2022})}\BibitemShut {NoStop}%
\bibitem [{\citenamefont {Schiller}\ \emph {et~al.}(2024)\citenamefont {Schiller}, \citenamefont {Alkalay}, \citenamefont {Hong}, \citenamefont {Umansky}, \citenamefont {Heiblum}, \citenamefont {Oreg},\ and\ \citenamefont {Snizhko}}]{Schiller_2024}%
  \BibitemOpen
  \bibfield  {author} {\bibinfo {author} {\bibfnamefont {Noam}\ \bibnamefont {Schiller}}, \bibinfo {author} {\bibfnamefont {Tomer}\ \bibnamefont {Alkalay}}, \bibinfo {author} {\bibfnamefont {Changki}\ \bibnamefont {Hong}}, \bibinfo {author} {\bibfnamefont {Vladimir}\ \bibnamefont {Umansky}}, \bibinfo {author} {\bibfnamefont {Moty}\ \bibnamefont {Heiblum}}, \bibinfo {author} {\bibfnamefont {Yuval}\ \bibnamefont {Oreg}}, \ and\ \bibinfo {author} {\bibfnamefont {Kyrylo}\ \bibnamefont {Snizhko}},\ }\href {https://arxiv.org/abs/2403.17097} {\enquote {\bibinfo {title} {Scaling tunnelling noise in the fractional quantum hall effect tells about renormalization and breakdown of chiral luttinger liquid},}\ } (\bibinfo {year} {2024}),\ \Eprint {http://arxiv.org/abs/2403.17097} {arXiv:2403.17097 [cond-mat.mes-hall]} \BibitemShut {NoStop}%
\bibitem [{\citenamefont {Tu}\ \emph {et~al.}(2013)\citenamefont {Tu}, \citenamefont {Zhang},\ and\ \citenamefont {Qi}}]{Tu_2013}%
  \BibitemOpen
  \bibfield  {author} {\bibinfo {author} {\bibfnamefont {Hong-Hao}\ \bibnamefont {Tu}}, \bibinfo {author} {\bibfnamefont {Yi}~\bibnamefont {Zhang}}, \ and\ \bibinfo {author} {\bibfnamefont {Xiao-Liang}\ \bibnamefont {Qi}},\ }\bibfield  {title} {\enquote {\bibinfo {title} {Momentum polarization: An entanglement measure of topological spin and chiral central charge},}\ }\href {\doibase 10.1103/PhysRevB.88.195412} {\bibfield  {journal} {\bibinfo  {journal} {Phys. Rev. B}\ }\textbf {\bibinfo {volume} {88}},\ \bibinfo {pages} {195412} (\bibinfo {year} {2013})}\BibitemShut {NoStop}%
\bibitem [{\citenamefont {Zaletel}\ \emph {et~al.}(2013)\citenamefont {Zaletel}, \citenamefont {Mong},\ and\ \citenamefont {Pollmann}}]{Zaletel_2013}%
  \BibitemOpen
  \bibfield  {author} {\bibinfo {author} {\bibfnamefont {M.~P.}\ \bibnamefont {Zaletel}}, \bibinfo {author} {\bibfnamefont {R.~S.~K.}\ \bibnamefont {Mong}}, \ and\ \bibinfo {author} {\bibfnamefont {F.}~\bibnamefont {Pollmann}},\ }\bibfield  {title} {\enquote {\bibinfo {title} {Topological characterization of fractional quantum hall ground states from microscopic hamiltonians},}\ }\href {\doibase 10.1103/PhysRevLett.110.236801} {\bibfield  {journal} {\bibinfo  {journal} {Phys. Rev. Lett.}\ }\textbf {\bibinfo {volume} {110}},\ \bibinfo {pages} {236801} (\bibinfo {year} {2013})}\BibitemShut {NoStop}%
\bibitem [{\citenamefont {Zaletel}\ and\ \citenamefont {Mong}(2012)}]{Zaletel_2012}%
  \BibitemOpen
  \bibfield  {author} {\bibinfo {author} {\bibfnamefont {M.~P.}\ \bibnamefont {Zaletel}}\ and\ \bibinfo {author} {\bibfnamefont {R.~S.~K.}\ \bibnamefont {Mong}},\ }\bibfield  {title} {\enquote {\bibinfo {title} {Exact matrix product states for quantum hall wave functions},}\ }\href {\doibase 10.1103/PhysRevB.86.245305} {\bibfield  {journal} {\bibinfo  {journal} {Phys. Rev. B}\ }\textbf {\bibinfo {volume} {86}},\ \bibinfo {pages} {245305} (\bibinfo {year} {2012})}\BibitemShut {NoStop}%
\bibitem [{\citenamefont {Read}\ and\ \citenamefont {Rezayi}(1999)}]{Read_1999}%
  \BibitemOpen
  \bibfield  {author} {\bibinfo {author} {\bibfnamefont {N.}~\bibnamefont {Read}}\ and\ \bibinfo {author} {\bibfnamefont {E.}~\bibnamefont {Rezayi}},\ }\bibfield  {title} {\enquote {\bibinfo {title} {Beyond paired quantum hall states: Parafermions and incompressible states in the first excited landau level},}\ }\href {\doibase 10.1103/PhysRevB.59.8084} {\bibfield  {journal} {\bibinfo  {journal} {Phys. Rev. B}\ }\textbf {\bibinfo {volume} {59}},\ \bibinfo {pages} {8084--8092} (\bibinfo {year} {1999})}\BibitemShut {NoStop}%
\bibitem [{\citenamefont {Fagerlund}\ and\ \citenamefont {Ardonne}(2024)}]{Fagerlund_2024}%
  \BibitemOpen
  \bibfield  {author} {\bibinfo {author} {\bibfnamefont {Alexander}\ \bibnamefont {Fagerlund}}\ and\ \bibinfo {author} {\bibfnamefont {Eddy}\ \bibnamefont {Ardonne}},\ }\href {https://arxiv.org/abs/2412.14889} {\enquote {\bibinfo {title} {A bosonic matrix product state description of read-rezayi states and its application to quasi-hole spins},}\ } (\bibinfo {year} {2024}),\ \Eprint {http://arxiv.org/abs/2412.14889} {arXiv:2412.14889 [cond-mat.str-el]} \BibitemShut {NoStop}%
\bibitem [{\citenamefont {Estienne}\ \emph {et~al.}(2013{\natexlab{a}})\citenamefont {Estienne}, \citenamefont {Papi\ifmmode~\acute{c}\else \'{c}\fi{}}, \citenamefont {Regnault},\ and\ \citenamefont {Bernevig}}]{Estienne_2013}%
  \BibitemOpen
  \bibfield  {author} {\bibinfo {author} {\bibfnamefont {B.}~\bibnamefont {Estienne}}, \bibinfo {author} {\bibfnamefont {Z.}~\bibnamefont {Papi\ifmmode~\acute{c}\else \'{c}\fi{}}}, \bibinfo {author} {\bibfnamefont {N.}~\bibnamefont {Regnault}}, \ and\ \bibinfo {author} {\bibfnamefont {B.~A.}\ \bibnamefont {Bernevig}},\ }\bibfield  {title} {\enquote {\bibinfo {title} {Matrix product states for trial quantum hall states},}\ }\href {\doibase 10.1103/PhysRevB.87.161112} {\bibfield  {journal} {\bibinfo  {journal} {Phys. Rev. B}\ }\textbf {\bibinfo {volume} {87}},\ \bibinfo {pages} {161112} (\bibinfo {year} {2013}{\natexlab{a}})}\BibitemShut {NoStop}%
\bibitem [{\citenamefont {Estienne}\ \emph {et~al.}(2013{\natexlab{b}})\citenamefont {Estienne}, \citenamefont {Regnault},\ and\ \citenamefont {Bernevig}}]{Estienne_2013b}%
  \BibitemOpen
  \bibfield  {author} {\bibinfo {author} {\bibfnamefont {B.}~\bibnamefont {Estienne}}, \bibinfo {author} {\bibfnamefont {N.}~\bibnamefont {Regnault}}, \ and\ \bibinfo {author} {\bibfnamefont {B.~A.}\ \bibnamefont {Bernevig}},\ }\bibfield  {title} {\enquote {\bibinfo {title} {Fractional quantum hall matrix product states for interacting conformal field theories},}\ }\href {\doibase 10.48550/arXiv.1311.2936} {\bibfield  {journal} {\bibinfo  {journal} {arXiv:1311.2936}\ } (\bibinfo {year} {2013}{\natexlab{b}}),\ 10.48550/arXiv.1311.2936}\BibitemShut {NoStop}%
\bibitem [{\citenamefont {Wu}\ \emph {et~al.}(2014)\citenamefont {Wu}, \citenamefont {Estienne}, \citenamefont {Regnault},\ and\ \citenamefont {Bernevig}}]{Wu_2014}%
  \BibitemOpen
  \bibfield  {author} {\bibinfo {author} {\bibfnamefont {Yang-Le}\ \bibnamefont {Wu}}, \bibinfo {author} {\bibfnamefont {B.}~\bibnamefont {Estienne}}, \bibinfo {author} {\bibfnamefont {N.}~\bibnamefont {Regnault}}, \ and\ \bibinfo {author} {\bibfnamefont {B.~Andrei}\ \bibnamefont {Bernevig}},\ }\bibfield  {title} {\enquote {\bibinfo {title} {Braiding non-abelian quasiholes in fractional quantum hall states},}\ }\href {\doibase 10.1103/PhysRevLett.113.116801} {\bibfield  {journal} {\bibinfo  {journal} {Phys. Rev. Lett.}\ }\textbf {\bibinfo {volume} {113}},\ \bibinfo {pages} {116801} (\bibinfo {year} {2014})}\BibitemShut {NoStop}%
\bibitem [{\citenamefont {Wu}\ \emph {et~al.}(2015)\citenamefont {Wu}, \citenamefont {Estienne}, \citenamefont {Regnault},\ and\ \citenamefont {Bernevig}}]{Wu2015MPS}%
  \BibitemOpen
  \bibfield  {author} {\bibinfo {author} {\bibfnamefont {Yang-Le}\ \bibnamefont {Wu}}, \bibinfo {author} {\bibfnamefont {B.}~\bibnamefont {Estienne}}, \bibinfo {author} {\bibfnamefont {N.}~\bibnamefont {Regnault}}, \ and\ \bibinfo {author} {\bibfnamefont {B.~Andrei}\ \bibnamefont {Bernevig}},\ }\bibfield  {title} {\enquote {\bibinfo {title} {Matrix product state representation of non-abelian quasiholes},}\ }\href {\doibase 10.1103/PhysRevB.92.045109} {\bibfield  {journal} {\bibinfo  {journal} {Phys. Rev. B}\ }\textbf {\bibinfo {volume} {92}},\ \bibinfo {pages} {045109} (\bibinfo {year} {2015})}\BibitemShut {NoStop}%
\bibitem [{\citenamefont {Herviou}\ and\ \citenamefont {Mila}(2024)}]{Herviou_2024}%
  \BibitemOpen
  \bibfield  {author} {\bibinfo {author} {\bibfnamefont {Lo\"{\i}c}\ \bibnamefont {Herviou}}\ and\ \bibinfo {author} {\bibfnamefont {Fr\'ed\'eric}\ \bibnamefont {Mila}},\ }\bibfield  {title} {\enquote {\bibinfo {title} {Numerical investigation of the structure factors of the read-rezayi series},}\ }\href {\doibase 10.1103/PhysRevB.110.045143} {\bibfield  {journal} {\bibinfo  {journal} {Phys. Rev. B}\ }\textbf {\bibinfo {volume} {110}},\ \bibinfo {pages} {045143} (\bibinfo {year} {2024})}\BibitemShut {NoStop}%
\bibitem [{\citenamefont {Thouless}(1984)}]{Thouless_1984}%
  \BibitemOpen
  \bibfield  {author} {\bibinfo {author} {\bibfnamefont {D.J.}\ \bibnamefont {Thouless}},\ }\bibfield  {title} {\enquote {\bibinfo {title} {Theory of the quantized hall effect},}\ }\href {\doibase https://doi.org/10.1016/0039-6028(84)90299-1} {\bibfield  {journal} {\bibinfo  {journal} {Surface Science}\ }\textbf {\bibinfo {volume} {142}},\ \bibinfo {pages} {147--154} (\bibinfo {year} {1984})}\BibitemShut {NoStop}%
\bibitem [{\citenamefont {Dubail}\ \emph {et~al.}(2012)\citenamefont {Dubail}, \citenamefont {Read},\ and\ \citenamefont {Rezayi}}]{Dubail_2012}%
  \BibitemOpen
  \bibfield  {author} {\bibinfo {author} {\bibfnamefont {J.}~\bibnamefont {Dubail}}, \bibinfo {author} {\bibfnamefont {N.}~\bibnamefont {Read}}, \ and\ \bibinfo {author} {\bibfnamefont {E.~H.}\ \bibnamefont {Rezayi}},\ }\bibfield  {title} {\enquote {\bibinfo {title} {Edge-state inner products and real-space entanglement spectrum of trial quantum hall states},}\ }\href {\doibase 10.1103/PhysRevB.86.245310} {\bibfield  {journal} {\bibinfo  {journal} {Phys. Rev. B}\ }\textbf {\bibinfo {volume} {86}},\ \bibinfo {pages} {245310} (\bibinfo {year} {2012})}\BibitemShut {NoStop}%
\bibitem [{\citenamefont {Li}(1993)}]{Li_1993}%
  \BibitemOpen
  \bibfield  {author} {\bibinfo {author} {\bibfnamefont {D.}~\bibnamefont {Li}},\ }\bibfield  {title} {\enquote {\bibinfo {title} {Intrinsic quasiparticle's spin and fractional quantum hall effect on riemann surfaces},}\ }\href {\doibase 10.1142/S0217984993001090} {\bibfield  {journal} {\bibinfo  {journal} {Mod. Phys. Lett. B}\ }\textbf {\bibinfo {volume} {07}},\ \bibinfo {pages} {1103--1110} (\bibinfo {year} {1993})}\BibitemShut {NoStop}%
\bibitem [{\citenamefont {Leinaas}(2002)}]{Leinaas2002}%
  \BibitemOpen
  \bibfield  {author} {\bibinfo {author} {\bibfnamefont {Jon~Magne}\ \bibnamefont {Leinaas}},\ }\enquote {\bibinfo {title} {Spin and statistics for quantum hall quasi-particles},}\ in\ \href {\doibase 10.1007/0-306-47094-2_15} {\emph {\bibinfo {booktitle} {Confluence of Cosmology, Massive Neutrinos, Elementary Particles, and Gravitation}}},\ \bibinfo {editor} {edited by\ \bibinfo {editor} {\bibfnamefont {Behram~N.}\ \bibnamefont {Kursunoglu}}, \bibinfo {editor} {\bibfnamefont {Stephan~L.}\ \bibnamefont {Mintz}}, \ and\ \bibinfo {editor} {\bibfnamefont {Arnold}\ \bibnamefont {Perlmutter}}}\ (\bibinfo  {publisher} {Springer US},\ \bibinfo {address} {Boston, MA},\ \bibinfo {year} {2002})\ pp.\ \bibinfo {pages} {149--161}\BibitemShut {NoStop}%
\bibitem [{\citenamefont {Read}(2009)}]{Read_2009}%
  \BibitemOpen
  \bibfield  {author} {\bibinfo {author} {\bibfnamefont {N.}~\bibnamefont {Read}},\ }\bibfield  {title} {\enquote {\bibinfo {title} {Non-abelian adiabatic statistics and hall viscosity in quantum hall states and $p_x+ip_y$ paired superfluids},}\ }\href {\doibase 10.1103/PhysRevB.79.045308} {\bibfield  {journal} {\bibinfo  {journal} {Phys. Rev. B}\ }\textbf {\bibinfo {volume} {79}},\ \bibinfo {pages} {045308} (\bibinfo {year} {2009})}\BibitemShut {NoStop}%
\bibitem [{\citenamefont {Gromov}(2016)}]{Gromov_2016}%
  \BibitemOpen
  \bibfield  {author} {\bibinfo {author} {\bibfnamefont {Andrey}\ \bibnamefont {Gromov}},\ }\bibfield  {title} {\enquote {\bibinfo {title} {Geometric defects in quantum hall states},}\ }\href {\doibase 10.1103/PhysRevB.94.085116} {\bibfield  {journal} {\bibinfo  {journal} {Phys. Rev. B}\ }\textbf {\bibinfo {volume} {94}},\ \bibinfo {pages} {085116} (\bibinfo {year} {2016})}\BibitemShut {NoStop}%
\bibitem [{\citenamefont {Umucal$\i$lar}\ \emph {et~al.}(2018)\citenamefont {Umucal$\i$lar}, \citenamefont {Macaluso}, \citenamefont {Comparin},\ and\ \citenamefont {Carusotto}}]{Umucalilar_2018}%
  \BibitemOpen
  \bibfield  {author} {\bibinfo {author} {\bibfnamefont {R.~O.}\ \bibnamefont {Umucal$\i$lar}}, \bibinfo {author} {\bibfnamefont {E.}~\bibnamefont {Macaluso}}, \bibinfo {author} {\bibfnamefont {T.}~\bibnamefont {Comparin}}, \ and\ \bibinfo {author} {\bibfnamefont {I.}~\bibnamefont {Carusotto}},\ }\bibfield  {title} {\enquote {\bibinfo {title} {Time-of-flight measurements as a possible method to observe anyonic statistics},}\ }\href {\doibase 10.1103/PhysRevLett.120.230403} {\bibfield  {journal} {\bibinfo  {journal} {Phys. Rev. Lett.}\ }\textbf {\bibinfo {volume} {120}},\ \bibinfo {pages} {230403} (\bibinfo {year} {2018})}\BibitemShut {NoStop}%
\bibitem [{\citenamefont {Macaluso}\ \emph {et~al.}(2019)\citenamefont {Macaluso}, \citenamefont {Comparin}, \citenamefont {Mazza},\ and\ \citenamefont {Carusotto}}]{Macaluso_2019}%
  \BibitemOpen
  \bibfield  {author} {\bibinfo {author} {\bibfnamefont {E.}~\bibnamefont {Macaluso}}, \bibinfo {author} {\bibfnamefont {T.}~\bibnamefont {Comparin}}, \bibinfo {author} {\bibfnamefont {L.}~\bibnamefont {Mazza}}, \ and\ \bibinfo {author} {\bibfnamefont {I.}~\bibnamefont {Carusotto}},\ }\bibfield  {title} {\enquote {\bibinfo {title} {Fusion channels of non-abelian anyons from angular-momentum and density-profile measurements},}\ }\href {\doibase 10.1103/PhysRevLett.123.266801} {\bibfield  {journal} {\bibinfo  {journal} {Phys. Rev. Lett.}\ }\textbf {\bibinfo {volume} {123}},\ \bibinfo {pages} {266801} (\bibinfo {year} {2019})}\BibitemShut {NoStop}%
\bibitem [{\citenamefont {Macaluso}\ \emph {et~al.}(2020)\citenamefont {Macaluso}, \citenamefont {Comparin}, \citenamefont {Umucal$\i$lar}, \citenamefont {Gerster}, \citenamefont {Montangero}, \citenamefont {Rizzi},\ and\ \citenamefont {Carusotto}}]{Macaluso_2020}%
  \BibitemOpen
  \bibfield  {author} {\bibinfo {author} {\bibfnamefont {E.}~\bibnamefont {Macaluso}}, \bibinfo {author} {\bibfnamefont {T.}~\bibnamefont {Comparin}}, \bibinfo {author} {\bibfnamefont {R.~O.}\ \bibnamefont {Umucal$\i$lar}}, \bibinfo {author} {\bibfnamefont {M.}~\bibnamefont {Gerster}}, \bibinfo {author} {\bibfnamefont {S.}~\bibnamefont {Montangero}}, \bibinfo {author} {\bibfnamefont {M.}~\bibnamefont {Rizzi}}, \ and\ \bibinfo {author} {\bibfnamefont {I.}~\bibnamefont {Carusotto}},\ }\bibfield  {title} {\enquote {\bibinfo {title} {Charge and statistics of lattice quasiholes from density measurements: A tree tensor network study},}\ }\href {\doibase 10.1103/PhysRevResearch.2.013145} {\bibfield  {journal} {\bibinfo  {journal} {Phys. Rev. Research}\ }\textbf {\bibinfo {volume} {2}},\ \bibinfo {pages} {013145} (\bibinfo {year} {2020})}\BibitemShut {NoStop}%
\bibitem [{Note1()}]{Note1}%
  \BibitemOpen
  \bibinfo {note} {We note that due to the screening on the FQH liquid, this relation holds up to corrections that are exponentially small in the distance to the quasiparticle.}\BibitemShut {Stop}%
\bibitem [{\citenamefont {Park}\ and\ \citenamefont {Haldane}(2014)}]{Park_2014}%
  \BibitemOpen
  \bibfield  {author} {\bibinfo {author} {\bibfnamefont {YeJe}\ \bibnamefont {Park}}\ and\ \bibinfo {author} {\bibfnamefont {F.~D.~M.}\ \bibnamefont {Haldane}},\ }\bibfield  {title} {\enquote {\bibinfo {title} {Guiding-center hall viscosity and intrinsic dipole moment along edges of incompressible fractional quantum hall fluids},}\ }\href {\doibase 10.1103/PhysRevB.90.045123} {\bibfield  {journal} {\bibinfo  {journal} {Phys. Rev. B}\ }\textbf {\bibinfo {volume} {90}},\ \bibinfo {pages} {045123} (\bibinfo {year} {2014})}\BibitemShut {NoStop}%
\bibitem [{\citenamefont {Wen}\ and\ \citenamefont {Zee}(1992)}]{Wen_1992}%
  \BibitemOpen
  \bibfield  {author} {\bibinfo {author} {\bibfnamefont {X.~G.}\ \bibnamefont {Wen}}\ and\ \bibinfo {author} {\bibfnamefont {A.}~\bibnamefont {Zee}},\ }\bibfield  {title} {\enquote {\bibinfo {title} {Shift and spin vector: New topological quantum numbers for the hall fluids},}\ }\href {\doibase 10.1103/PhysRevLett.69.953} {\bibfield  {journal} {\bibinfo  {journal} {Phys. Rev. Lett.}\ }\textbf {\bibinfo {volume} {69}},\ \bibinfo {pages} {953--956} (\bibinfo {year} {1992})}\BibitemShut {NoStop}%
\bibitem [{\citenamefont {Haldane}(2009)}]{haldane2009hallviscosityintrinsicmetric}%
  \BibitemOpen
  \bibfield  {author} {\bibinfo {author} {\bibfnamefont {F.~D.~M.}\ \bibnamefont {Haldane}},\ }\href {https://arxiv.org/abs/0906.1854} {\enquote {\bibinfo {title} {"hall viscosity" and intrinsic metric of incompressible fractional hall fluids},}\ } (\bibinfo {year} {2009}),\ \Eprint {http://arxiv.org/abs/0906.1854} {arXiv:0906.1854 [cond-mat.str-el]} \BibitemShut {NoStop}%
\bibitem [{\citenamefont {Simon}(2021)}]{wavefunctionology}%
  \BibitemOpen
  \bibfield  {author} {\bibinfo {author} {\bibfnamefont {Steven~H.}\ \bibnamefont {Simon}},\ }\href {https://arxiv.org/abs/2107.00437} {\enquote {\bibinfo {title} {Wavefunctionology: The special structure of certain fractional quantum hall wavefunctions},}\ } (\bibinfo {year} {2021}),\ \Eprint {http://arxiv.org/abs/2107.00437} {arXiv:2107.00437 [cond-mat.str-el]} \BibitemShut {NoStop}%
\bibitem [{Note2()}]{Note2}%
  \BibitemOpen
  \bibinfo {note} {Conservation of charge implies that one can, in principle, define $\protect \bar {x}_0$ uniquely even in actual physical situations, where the edge structure depends on details of the experimental setup}\BibitemShut {NoStop}%
\bibitem [{\citenamefont {Thouless}\ and\ \citenamefont {Wu}(1985)}]{Thouless_1985}%
  \BibitemOpen
  \bibfield  {author} {\bibinfo {author} {\bibfnamefont {D.~J.}\ \bibnamefont {Thouless}}\ and\ \bibinfo {author} {\bibfnamefont {Y.-S.}\ \bibnamefont {Wu}},\ }\bibfield  {title} {\enquote {\bibinfo {title} {Remarks on fractional statistics},}\ }\href {\doibase 10.1103/PhysRevB.31.1191} {\bibfield  {journal} {\bibinfo  {journal} {Phys. Rev. B}\ }\textbf {\bibinfo {volume} {31}},\ \bibinfo {pages} {1191--1193} (\bibinfo {year} {1985})}\BibitemShut {NoStop}%
\bibitem [{\citenamefont {Preskill}(2004)}]{Preskill_2004}%
  \BibitemOpen
  \bibfield  {author} {\bibinfo {author} {\bibfnamefont {J.}~\bibnamefont {Preskill}},\ }\href {http://www.theory.caltech.edu/~preskill/ph219/topological.pdf} {\emph {\bibinfo {title} {Lecture notes Lecture Notes for Physics 219: Quantum Computation}}}\ (\bibinfo {year} {2004})\ Chap.\ \bibinfo {chapter} {9: Topological quantum computation}\BibitemShut {NoStop}%
\bibitem [{\citenamefont {Cappelli}\ \emph {et~al.}(2001)\citenamefont {Cappelli}, \citenamefont {Georgiev},\ and\ \citenamefont {Todorov}}]{Cappelli_2001}%
  \BibitemOpen
  \bibfield  {author} {\bibinfo {author} {\bibfnamefont {Andrea}\ \bibnamefont {Cappelli}}, \bibinfo {author} {\bibfnamefont {Lachezar~S.}\ \bibnamefont {Georgiev}}, \ and\ \bibinfo {author} {\bibfnamefont {Ivan~T.}\ \bibnamefont {Todorov}},\ }\bibfield  {title} {\enquote {\bibinfo {title} {Parafermion hall states from coset projections of abelian conformal theories},}\ }\href {\doibase https://doi.org/10.1016/S0550-3213(00)00774-4} {\bibfield  {journal} {\bibinfo  {journal} {Nuclear Physics B}\ }\textbf {\bibinfo {volume} {599}},\ \bibinfo {pages} {499--530} (\bibinfo {year} {2001})}\BibitemShut {NoStop}%
\bibitem [{\citenamefont {Ardonne}\ and\ \citenamefont {Schoutens}(2007)}]{Ardonne_2007}%
  \BibitemOpen
  \bibfield  {author} {\bibinfo {author} {\bibfnamefont {E.}~\bibnamefont {Ardonne}}\ and\ \bibinfo {author} {\bibfnamefont {K.}~\bibnamefont {Schoutens}},\ }\bibfield  {title} {\enquote {\bibinfo {title} {Wavefunctions for topological quantum registers},}\ }\href {\doibase https://doi.org/10.1016/j.aop.2006.07.015} {\bibfield  {journal} {\bibinfo  {journal} {Annals of Physics}\ }\textbf {\bibinfo {volume} {322}},\ \bibinfo {pages} {201--235} (\bibinfo {year} {2007})},\ \bibinfo {note} {january Special Issue 2007}\BibitemShut {NoStop}%
\bibitem [{\citenamefont {Cr\'epel}\ \emph {et~al.}(2018)\citenamefont {Cr\'epel}, \citenamefont {Estienne}, \citenamefont {Bernevig}, \citenamefont {Lecheminant},\ and\ \citenamefont {Regnault}}]{Crepel_2018}%
  \BibitemOpen
  \bibfield  {author} {\bibinfo {author} {\bibfnamefont {V.}~\bibnamefont {Cr\'epel}}, \bibinfo {author} {\bibfnamefont {B.}~\bibnamefont {Estienne}}, \bibinfo {author} {\bibfnamefont {B.~A.}\ \bibnamefont {Bernevig}}, \bibinfo {author} {\bibfnamefont {P.}~\bibnamefont {Lecheminant}}, \ and\ \bibinfo {author} {\bibfnamefont {N.}~\bibnamefont {Regnault}},\ }\bibfield  {title} {\enquote {\bibinfo {title} {Matrix product state description of halperin states},}\ }\href {\doibase 10.1103/PhysRevB.97.165136} {\bibfield  {journal} {\bibinfo  {journal} {Phys. Rev. B}\ }\textbf {\bibinfo {volume} {97}},\ \bibinfo {pages} {165136} (\bibinfo {year} {2018})}\BibitemShut {NoStop}%
\bibitem [{\citenamefont {Cr\'epel}\ \emph {et~al.}(2019)\citenamefont {Cr\'epel}, \citenamefont {Regnault},\ and\ \citenamefont {Estienne}}]{Crepel_2019}%
  \BibitemOpen
  \bibfield  {author} {\bibinfo {author} {\bibfnamefont {Valentin}\ \bibnamefont {Cr\'epel}}, \bibinfo {author} {\bibfnamefont {Nicolas}\ \bibnamefont {Regnault}}, \ and\ \bibinfo {author} {\bibfnamefont {Benoit}\ \bibnamefont {Estienne}},\ }\bibfield  {title} {\enquote {\bibinfo {title} {Matrix product state description and gaplessness of the haldane-rezayi state},}\ }\href {\doibase 10.1103/PhysRevB.100.125128} {\bibfield  {journal} {\bibinfo  {journal} {Phys. Rev. B}\ }\textbf {\bibinfo {volume} {100}},\ \bibinfo {pages} {125128} (\bibinfo {year} {2019})}\BibitemShut {NoStop}%
\bibitem [{\citenamefont {Halperin}(1983)}]{Halperin_1983}%
  \BibitemOpen
  \bibfield  {author} {\bibinfo {author} {\bibfnamefont {B.~I.}\ \bibnamefont {Halperin}},\ }\bibfield  {title} {\enquote {\bibinfo {title} {Theory of the quantized hall conductance},}\ }\href {https://www.e-periodica.ch/digbib/view?pid=hpa-001:1983:56::1243#87} {\bibfield  {journal} {\bibinfo  {journal} {Helv. Phys. Acta}\ }\textbf {\bibinfo {volume} {56}},\ \bibinfo {pages} {75--102} (\bibinfo {year} {1983})}\BibitemShut {NoStop}%
\bibitem [{\citenamefont {Haldane}\ and\ \citenamefont {Rezayi}(1988)}]{Haldane_1988}%
  \BibitemOpen
  \bibfield  {author} {\bibinfo {author} {\bibfnamefont {F.~D.~M.}\ \bibnamefont {Haldane}}\ and\ \bibinfo {author} {\bibfnamefont {E.~H.}\ \bibnamefont {Rezayi}},\ }\bibfield  {title} {\enquote {\bibinfo {title} {Spin-singlet wave function for the half-integral quantum hall effect},}\ }\href {\doibase 10.1103/PhysRevLett.60.956} {\bibfield  {journal} {\bibinfo  {journal} {Phys. Rev. Lett.}\ }\textbf {\bibinfo {volume} {60}},\ \bibinfo {pages} {956--959} (\bibinfo {year} {1988})}\BibitemShut {NoStop}%
\bibitem [{\citenamefont {Haldane}(2011)}]{Haldane_2011}%
  \BibitemOpen
  \bibfield  {author} {\bibinfo {author} {\bibfnamefont {F.~D.~M.}\ \bibnamefont {Haldane}},\ }\bibfield  {title} {\enquote {\bibinfo {title} {Geometrical description of the fractional quantum hall effect},}\ }\href {\doibase 10.1103/PhysRevLett.107.116801} {\bibfield  {journal} {\bibinfo  {journal} {Phys. Rev. Lett.}\ }\textbf {\bibinfo {volume} {107}},\ \bibinfo {pages} {116801} (\bibinfo {year} {2011})}\BibitemShut {NoStop}%
\bibitem [{\citenamefont {Haldane}\ and\ \citenamefont {Shen}(2016)}]{haldane2016geometrylandauorbitsabsence}%
  \BibitemOpen
  \bibfield  {author} {\bibinfo {author} {\bibfnamefont {F.~D.~M.}\ \bibnamefont {Haldane}}\ and\ \bibinfo {author} {\bibfnamefont {Yu}~\bibnamefont {Shen}},\ }\href {https://arxiv.org/abs/1512.04502} {\enquote {\bibinfo {title} {Geometry of landau orbits in the absence of rotational symmetry},}\ } (\bibinfo {year} {2016}),\ \Eprint {http://arxiv.org/abs/1512.04502} {arXiv:1512.04502 [cond-mat.mes-hall]} \BibitemShut {NoStop}%
\bibitem [{\citenamefont {Haldane}(2023)}]{haldane2023incompressiblequantumhallfluids}%
  \BibitemOpen
  \bibfield  {author} {\bibinfo {author} {\bibfnamefont {F.~D.~M.}\ \bibnamefont {Haldane}},\ }\href {https://arxiv.org/abs/2302.12472} {\enquote {\bibinfo {title} {Incompressible quantum hall fluids as electric quadrupole fluids},}\ } (\bibinfo {year} {2023}),\ \Eprint {http://arxiv.org/abs/2302.12472} {arXiv:2302.12472 [cond-mat.str-el]} \BibitemShut {NoStop}%
\bibitem [{\citenamefont {Oblak}\ \emph {et~al.}(2024)\citenamefont {Oblak}, \citenamefont {Lapierre}, \citenamefont {Moosavi}, \citenamefont {St\'ephan},\ and\ \citenamefont {Estienne}}]{Oblak_2024}%
  \BibitemOpen
  \bibfield  {author} {\bibinfo {author} {\bibfnamefont {Blagoje}\ \bibnamefont {Oblak}}, \bibinfo {author} {\bibfnamefont {Bastien}\ \bibnamefont {Lapierre}}, \bibinfo {author} {\bibfnamefont {Per}\ \bibnamefont {Moosavi}}, \bibinfo {author} {\bibfnamefont {Jean-Marie}\ \bibnamefont {St\'ephan}}, \ and\ \bibinfo {author} {\bibfnamefont {Benoit}\ \bibnamefont {Estienne}},\ }\bibfield  {title} {\enquote {\bibinfo {title} {Anisotropic quantum hall droplets},}\ }\href {\doibase 10.1103/PhysRevX.14.011030} {\bibfield  {journal} {\bibinfo  {journal} {Phys. Rev. X}\ }\textbf {\bibinfo {volume} {14}},\ \bibinfo {pages} {011030} (\bibinfo {year} {2024})}\BibitemShut {NoStop}%
\end{thebibliography}%

\end{document}